\documentclass[sn-mathphys,Numbered]{sn-jnl}

\usepackage{xcolor}
\usepackage{graphicx}
\usepackage{amsthm}
\usepackage{amsmath}
\usepackage{amssymb}
\usepackage{algorithm}
\usepackage{algpseudocode}
\usepackage[numbers,sort&compress]{natbib}
\usepackage{soul}
\usepackage[percent]{overpic}
\usepackage[T1]{fontenc}
\usepackage[utf8]{inputenc}
\usepackage{caption}
\usepackage{xcolor}
\usepackage[normalem]{ulem}
\usepackage{etoc}

\newcommand{\new}[1]{
  {
   #1
  }
}

\theoremstyle{definition}
\newtheorem{definition}{Definition}[section]

\begin{document}

\title[Direction-aware TDA]{Direction-aware topological descriptors for elastic stiffness tensor prediction in porous materials}

\author*[1,2]{Rafał Topolnicki}

\author[1]{Michał Bogdan}
\author[3]{Jakub Malinowski}
\author[1,4]{Bartosz Naskręcki}
\author[5]{Maciej Harańczyk}
\author[1]{Paweł Dłotko}

\affil[1]{Dioscuri Center in Topological Data Analysis, Institute of Mathematics, Polish Academy of Sciences, ul.
Śniadeckich 8, 00-656 Warsaw, Poland}
\affil[2]{Institute of Experimental Physics, University of Wrocław, pl. Maxa Borna 9, Wrocław 50-204, Poland}
\affil[3]{Faculty of Pure and Applied Mathematics, Wrocław University of Science and Technology, ul. Wybrzeże Wyspiańskiego 27, 50-370 Wrocław, Poland}
\affil[4]{Faculty of Mathematics and Computer Science, Adam Mickiewicz University, ul. Uniwersytetu Poznańskiego
4, 61-614 Poznań, Poland}
\affil[5]{IMDEA Materials Institute, C. Eric Kandel 2, Getafe, 28906 Madrid, Spain}

\abstract{
Classical topological descriptors used in topological data analysis (TDA) are invariant under permutations of spatial axes and
therefore cannot represent the loading direction, which is essential for modeling anisotropic mechanical response. Here, this
limitation is addressed by introducing a \emph{direction-aware TDA framework} in which the loading axis is explicitly embedded into
filtration functions used to compute both persistent homology and Euler characteristic profile descriptors. 
We apply this framework to predict the full elastic stiffness tensor of porous microstructures using non-directional as well as direction-aware descriptors of the structures as well as convolutional neural networks  trained directly on the voxelized structure. We show that the performance of all those are comparable on the diagonal uniaxial, Poisson, and shear components. However for the twelve off-diagonal, normal shear coupling components – which govern elastic anisotropy and are the hardest to predict – only direction-aware topology retains meaningful predictive power, with all baselines, non-directional descriptors and the CNN  collapsing to near-chance accuracy. When used as inputs to gradient-boosted tree models, the proposed descriptors match or exceed the accuracy of the convolutional neural network specifically on these hardest-to-predict coupling terms, despite relying on a compact, physically interpretable representation that is orders of magnitude smaller than the raw voxel grid. Overall, the results establish direction-aware TDA as a practical route for linking porous microstructure to the full anisotropic elastic response, capturing coupling terms that conventional descriptors and end-to-end deep learning models fail to resolve.}

\maketitle

\section{Introduction}
Porous materials consist of interconnected regions of solid matter and voids, which may be filled with a gas or liquid.  Porous and nanoporous metals, in particular, have found widespread applications in chemical catalysis, plasmonics, and spectroscopy~\cite{Koya2021}, as orthopedic implants~\cite{Guo2025,Matassi2013,Depboylu2024}, in energy storage technologies, and as the basis for strong, ultralight structural materials~\cite{Schaedler2016}. In many of these applications, the mechanical performance of porous metals—encompassing stiffness, strength, deformation behavior, and failure—plays a critical role in determining functionality, durability, and reliability.

Despite decades of research, establishing predictive relationships between porous structure and mechanical properties remains challenging. In particular, predicting the dependence of Young’s modulus on microstructural features such as porosity, pore morphology, and connectivity is nontrivial. Most existing approaches rely on scaling relationships between material density and tensile or compressive moduli, derived from empirical observations or simplified theoretical models.
The widely used and theoretically well-justified Gibson-Ashby model~\cite{Gibson1982} postulates a power-law relationship between material density $\rho$ and Young's modulus $E \propto \rho^n$. This foundational model predicts $E \propto \rho^2$ for open-cell porous materials and $E \propto \rho^3 $ for materials with closed pores~\cite{Gibson1982}.
However, many types of materials violate the Gibson-Ashby model. In some open-celled nanoporous metal structures Young's modulus grows faster with density than the model's predictions suggest~\cite{Badwe2017}, while in others it grows slower~\cite{Zheng2014}. While this makes it clear that the porosity of a porous material itself does not determine its Young's modulus and geometrical and topological details of the solid material and pores also play key roles, no general functional relationship between Young's modulus and topology is known.

Topological data analysis (TDA) offers a principled mathematical framework for quantifying such structural features across multiple length scales. By characterizing connectivity, loops, and cavities in a scale-dependent manner, TDA provides compact summaries of complex structures that are robust to noise and discretization. Tools such as persistent homology, persistence diagrams, Euler characteristic curves, and Euler characteristic profiles encode how topological features emerge and disappear as a structure is coarse-grained. Importantly, these descriptors can be efficiently computed for large datasets and subsequently vectorized for use in machine-learning pipelines~\cite{Krishnapriyan2021,Moon2019}.

A limited number of such TDA-involving attempts to predict the properties of porous materials have been reported. The existing results suggest TDA can be helpful in quantifying the heterogeneity of porous materials and predicting properties related to flow transport, permeability, fluid trapping and dissolution properties under reactive transport~\cite{Robins2016, Herring2019, Moon2019, Thompson2023, Lisitsa2020}, classifying  metal-organic frameworks based on their gas adsorption potential~\cite{Lee2017} and predicting the efficiency of the adsorption~\cite{Krishnapriyan2021, Chen2025, Wang2025, Lee2017}. A relationship has been found between persistent homology and elastic modulus in a sample of rock structures ~\cite{Jiang2018} and a similar study was conducted for wet compact powders~\cite{Ishihara2023}, however, both of these studies are affected by a very small sample sizes.

A key limitation of existing TDA approaches in this context is their inherent isotropy. Standard topological descriptors are invariant under rotations and reflections and therefore do not encode directional information or account for the breaking of symmetry with respect to the principal axes. This poses a fundamental obstacle for modeling anisotropic mechanical response in porous materials, where uniaxial Young’s modulus depends strongly on the loading direction and different axes may act as mechanically easy or hard directions. As a result, direction-agnostic topological summaries cannot distinguish between structurally identical but mechanically inequivalent orientations, making them unsuitable for predicting direction-dependent properties such as uniaxial stiffness or strength in anisotropic porous solids.

In this work, a direction-aware TDA framework is introduced by embedding the loading axis directly into filtration functions and systematically evaluating the resulting descriptors for predicting the full elastic stiffness tensor -- from the uniaxial and Poisson-type normal components to the shear and normal-shear coupling (off-diagonal) terms. To this end, we construct datasets spanning a wide range of structural anisotropy and demonstrate that directional TDA descriptors achieve substantially higher predictive performance than the non-directional descriptors commonly used in the literature. This advantage is not distributed uniformly across the stiffness tensor: it is concentrated in the off-diagonal, normal-shear coupling terms, which conventional directional descriptors -- and even a convolutional neural network trained directly on the voxelized structure -- largely fail to capture.
For strongly anisotropic structures, direction-aware descriptors provide large gains in both $R^2$ (coefficient of determination) and MAE (Mean Absolute Error), while for nominally isotropic datasets they remain uniformly competitive and typically improve $R^2$, indicating that direction-aware topology captures mechanically relevant organization even when macroscopic anisotropy is weak.
To verify that these findings are not specific to a particular structure class or generation mechanism, we further evaluate the approach on additional datasets encompassing diverse porous topologies and controlled anisotropy levels, recovering the same qualitative trends. Across all datasets, direction-aware persistent homology and multifiltration Euler characteristic profiles yield superior predictive performance, with the performance gap increasing systematically with structural anisotropy. When combined, directional PH and ECP descriptors achieve predictive accuracy approaching that of convolutional neural networks trained directly on voxelized structure data, while remaining orders of magnitude more compact and offering substantially greater interpretability.

This paper is organized as follows. Section~2 describes the datasets of porous structures used in this study, including the procedures for dataset generation and the numerical estimation of uniaxial Young’s modulus using FFT-based simulations. Section~3 introduces the construction of both directional and non-directional topological descriptors, detailing the formulation of force-direction-aware scalar and vector fields for persistent homology and Euler characteristic profiles. 
Section~4 presents and analyzes the results of the proposed direction-aware TDA-based predictive models for the full stiffness
tensor, compares their performance with non-topological directional baselines (porosity, two-point correlation, fabric tensor),
the direction-agnostic TDA descriptor, and a convolutional neural network trained directly on voxelized structures, and reports
paired Wilcoxon signed-rank tests confirming that the advantage of direction-aware topology is concentrated in the off-diagonal
coupling terms of the elastic tensor.
Section~5 summarizes the main findings and discusses their implications.

\section{Dataset}

\subsection{Random Trigonometric Phase}

The porous two-phase structures used in this study were generated using a
\emph{random trigonometric phase} (RTP) approach, in which a continuous,
statistically homogeneous random field is constructed as a superposition of
trigonometric modes with random phases and wavevectors.

We consider a cubic domain discretized on a regular
$N \times N \times N$ grid with $N=80$ and periodic boundary conditions.
The spatial coordinates are normalized to the unit cube,
$\mathbf{x} \in [0,1)^3$, and periodicity is enforced implicitly by the
trigonometric basis.
The RTP scalar field $S(\mathbf{x})$ is defined as
\begin{equation}
\label{eqn:rtp_S}
S(\mathbf{x}) =
\sqrt{\frac{2}{K}}
\sum_{k=1}^{K}
\cos\!\left( 2\pi\,\mathbf{q}_k \cdot \mathbf{x} + \phi_k \right),
\end{equation}
where $K$ is the number of trigonometric modes,
$\phi_k \sim \mathcal{U}(0,2\pi)$ are independent random phase shifts, and the
wavevectors $\mathbf{q}_k$ are sampled from a bounded integer lattice $\mathbf{q}_k = (n_x, n_y, n_z), n_\alpha \in \{-n_\text{max}, -n_\text{max}+1,\ldots, n_\text{max}-1,n_\text{max}\}.$
Because each cosine mode is periodic on the unit torus, the field $S$ and all
derived microstructures satisfy periodic boundary conditions by construction.

The continuous field $S(\mathbf{x})$ is converted into a binary porous medium
by thresholding.
The material indicator function $X(\mathbf{x}) \in \{0,1\}$ is defined as
\[
X(\mathbf{x}) =
\begin{cases}
1, & S(\mathbf{x}) > \tau, \\
0, & \text{otherwise},
\end{cases}
\]
where $X=1$ denotes solid material and $X=0$ denotes void.
For each realization, the threshold $\tau$ is determined numerically such that
the resulting porosity $\phi$ matches a prescribed target value, selected
prior to generation.
To ensure physically meaningful structures, each realization is subjected to a
connectivity and percolation analysis under periodic boundary conditions:
all material voxels must form a single connected component and the solid phase
must percolate across the domain in all three Cartesian directions, otherwise the structure is not included in the dataset.
In this study, we used $K$ values sampled uniformly from the interval $[10, 30]$, fixed $n_{\text{max}} = 12$, and selected the target porosity uniformly from the interval $[0.2, 0.8]$.

In the isotropic case, the statistical properties of $S$ are invariant under
rotations, as the spectral content is identical on average along all spatial
directions.
Directional anisotropy is introduced explicitly in Fourier space by scaling
the components of the wavevectors prior to evaluating Eq.~\eqref{eqn:rtp_S}.
Specifically, integer wavevectors $\mathbf{q}_k^{(0)}$ are first sampled
uniformly from the bounded lattice, and the final wavevectors are defined as
\begin{equation*}
\mathbf{q}_k = \mathrm{diag}(s_x, s_y, s_z)\,\mathbf{q}_k^{(0)},
\end{equation*}
where $(s_x, s_y, s_z)$ are prescribed anisotropy scaling factors.
This construction directly controls the characteristic correlation lengths of
the random field: larger values of $s_\alpha$ correspond to higher spatial
frequencies and thus shorter correlation lengths along direction $\alpha$,
whereas smaller values of $s_\alpha$ lead to longer-range correlations.
In this work, we set $s_x = s_y = 1$ and $s_z = 0.2$,
with significantly extended correlation length along the $z$ axis and a
pronounced directional anisotropy.

\begin{figure}
    \centering
    \begin{overpic}[width=0.3\textwidth]{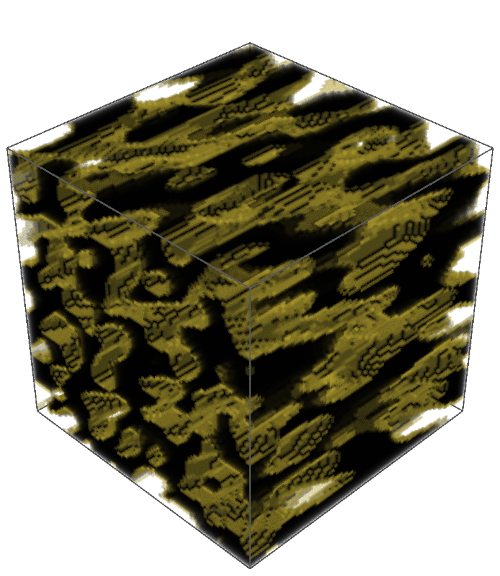}
        \put(5,85){\large a}
    \end{overpic}
    \begin{overpic}[width=0.3\textwidth]{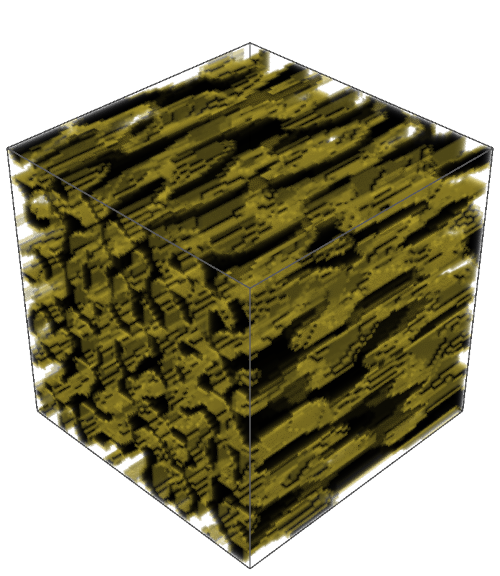}
        \put(5,85){\large b}
    \end{overpic}
    \begin{overpic}[width=0.3\textwidth]{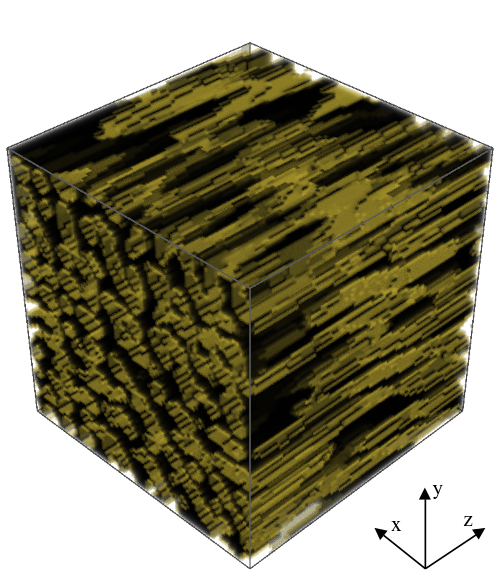}
        \put(5,85){\large c}
    \end{overpic}
    \caption{
    Three exemplary RTP structures spanning a range of porosities and directional mechanical responses. The
directional Young's moduli $E=(E_x,E_y,E_z)$ are: (a) $(10.3, 11.3, 31.4)\,\mathrm{GPa}$ at porosity
$0.60$, (b) $(5.2, 10.0, 33.1)\,\mathrm{GPa}$ at porosity $0.54$, and (c) $(19.8, 17.5,
51.2)\,\mathrm{GPa}$ at porosity $0.68$. In all three structures $E_z$ exceeds $E_x$ and $E_y$ by a wide
margin, consistent with the visible elongation of the pore structure along the $z$ direction and the
correspondingly stiffer mechanical response along that axis.
        }
    \label{fig:rtp}
\end{figure}

Figure~\ref{fig:rtp} presents three exemplary RTP structures spanning a range of structural anisotropy,
together with their directional Young's moduli $E_x, E_y, E_z$ -- the uniaxial diagonal components of the
stiffness tensor -- and porosities, illustrating correlation between directional morphology and mechanical
response.

\subsection{Dataset of topologically diverse structures}
\label{sec:td_structures}
\begin{figure}
    \centering
    \begin{overpic}[width=0.18\textwidth]{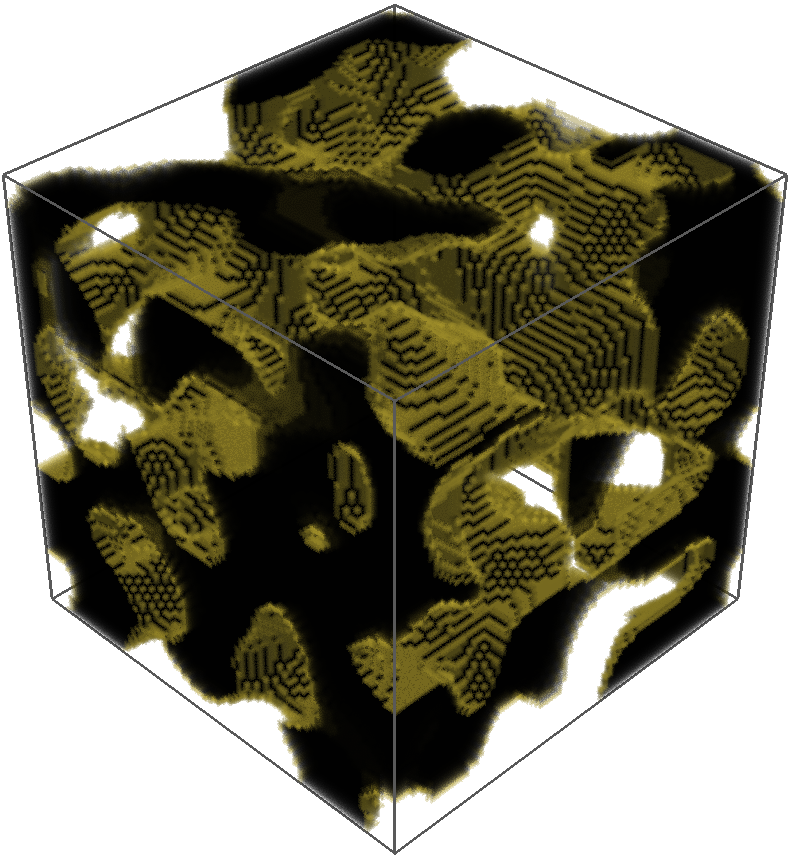}
        \put(5,85){\large a}
    \end{overpic}
    \begin{overpic}[width=0.18\textwidth]{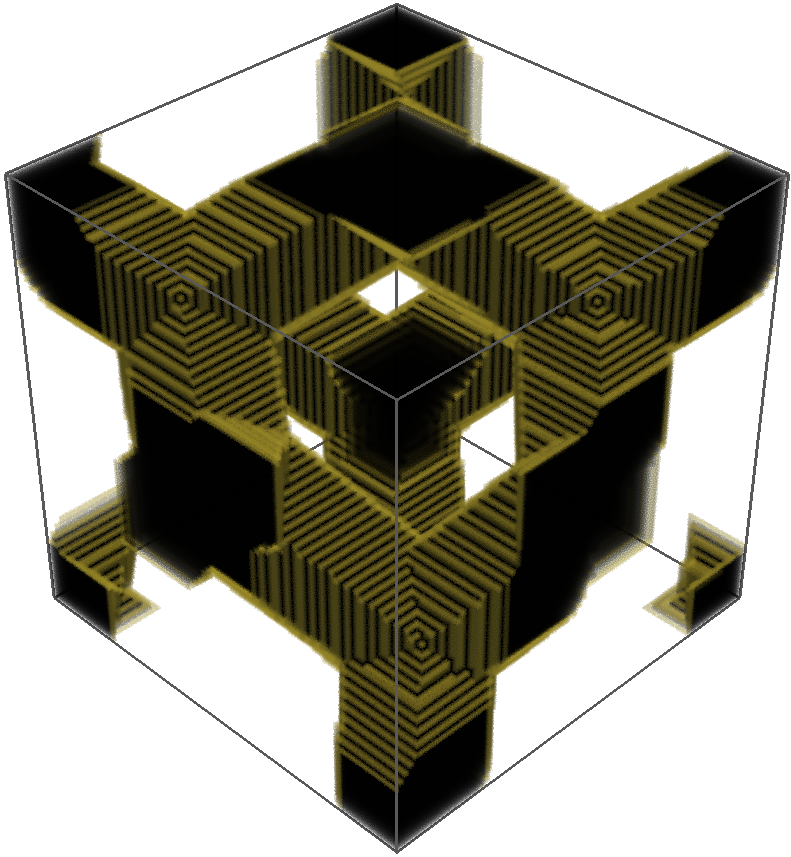}
        \put(5,85){\large b}
    \end{overpic}
    \begin{overpic}[width=0.18\textwidth]{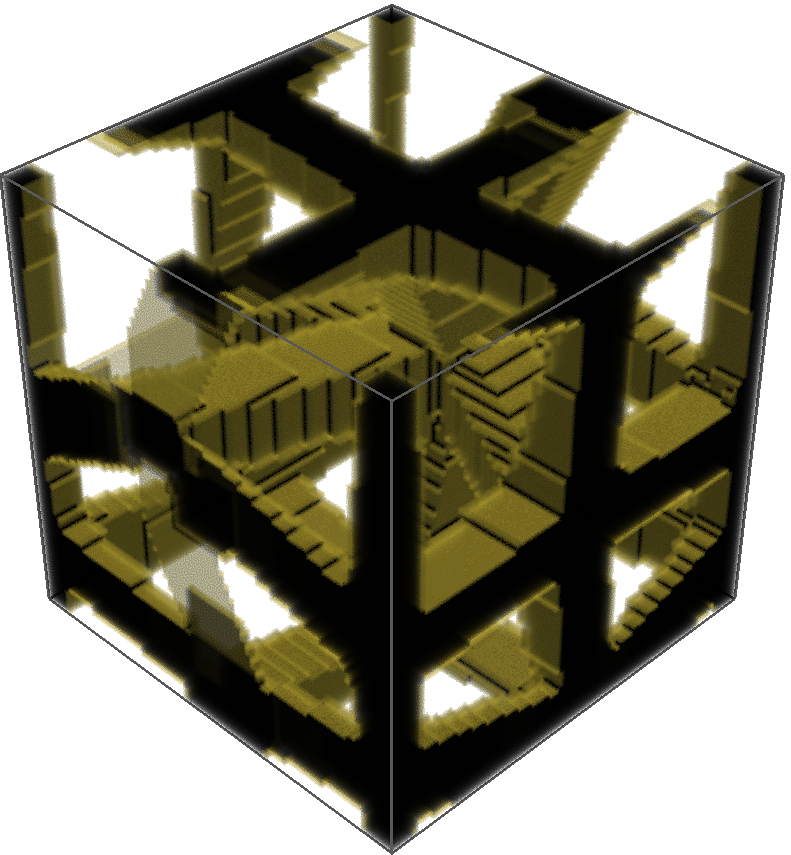}
        \put(5,85){\large c}
    \end{overpic}
    \begin{overpic}[width=0.18\textwidth]{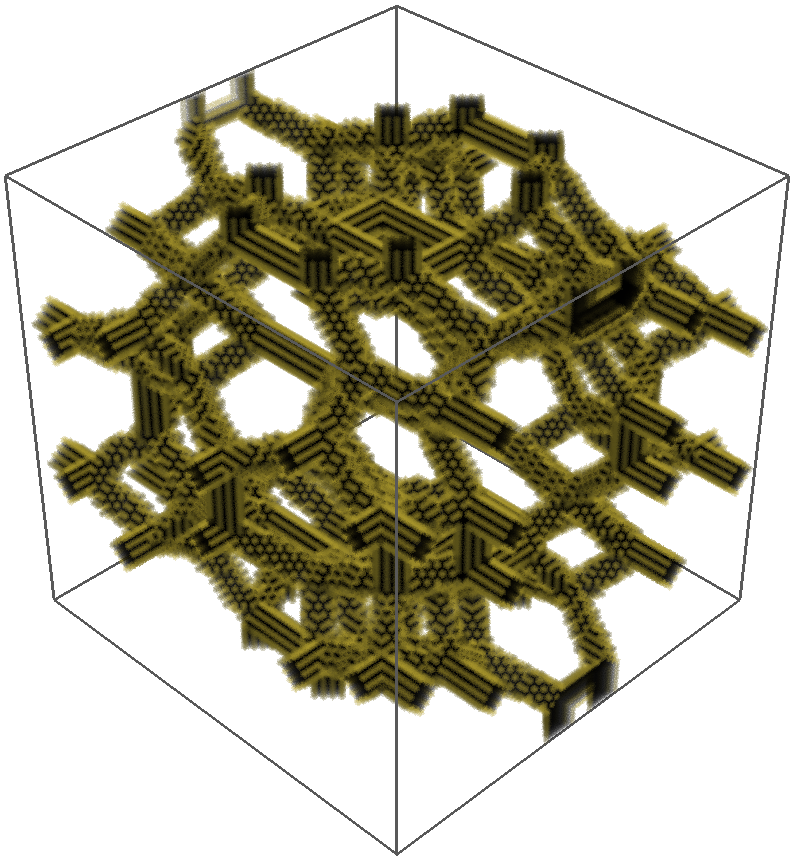}
        \put(5,85){\large d}
    \end{overpic}
    \begin{overpic}[width=0.18\textwidth]{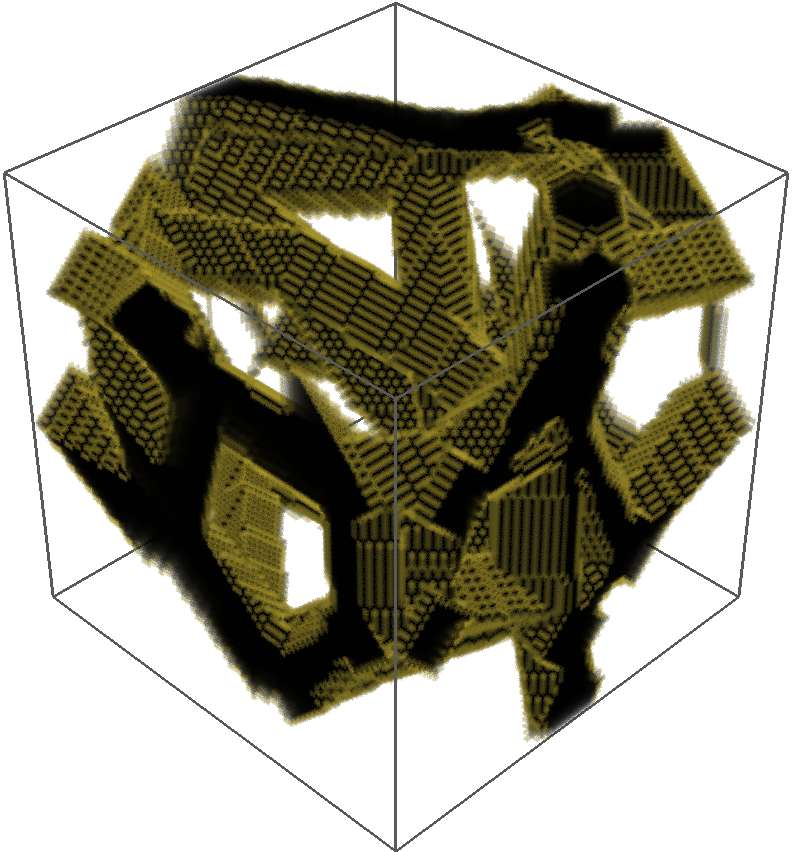}
        \put(5,85){\large e}
    \end{overpic}

    \caption{Examples of the five structure types used in the TD dataset: a: splines, b: diamond, c: cubic, d: zeolite, e: Voronoi-based structures.}
    \label{fig:various_aniso}
\end{figure}

A dataset of 2376 topologically diverse (TD)
periodic porous structures was generated to assess the
generality of the proposed topological descriptors beyond RTP-based structures.
All structures occupy a normalized $1\times1\times1$ cubic domain, voxelized on a
regular $80\times80\times80$ grid, and satisfy periodic boundary conditions in all three spatial
directions.
Importantly, the dataset was constructed to be \emph{statistically isotropic}:
no preferred spatial direction is imposed by the generation procedures, and the
resulting ensembles exhibit no systematic directional bias in their geometric
or mechanical properties.
Within this constraint, the dataset spans a broad range of pore morphologies and
topologies.
The dataset comprises five approximately equally sized subsets, each
corresponding to a distinct family of porous structures generated by a
different stochastic algorithm.
Within each family, randomness ensures substantial variability in geometric
features and effective elastic properties while preserving statistical
isotropy. Additional details on the generating algorithms are available in the SI, while the full dataset, together with the corresponding unidirectional Young’s moduli, is available in the dedicated repository associated with this paper -- see section Code and Data Availability for more information.
The five structure families are referred to as Voronoi, Zeolitic, Diamond-like,
Cubic-strut, and Spline-based structures.
Representative examples and generation schematics are shown in
Figure~\ref{fig:various_aniso}.
Below we briefly summarize the defining characteristics of each family.

\paragraph{Voronoi-based structures} were generated from three-dimensional Voronoi
tessellations of randomly distributed seed points within the unit cell.
Between three and twelve points were sampled uniformly per realization, and the
edges of the resulting tessellation were used as the structural skeleton.
These edges were voxelized and thickened to form struts with square cross
sections of randomly varying thickness (see Figure~S1 in the Supplementary Information).

\paragraph{Zeolitic-inspired structures} were derived from predicted zeolite frameworks reported in
the database of Deem and co-workers~\cite{C0CP02255A,deem2023pcod}.
A subset of structures with orthogonal unit cells was selected and rescaled to
the normalized cubic domain.
Although individual realizations may exhibit complex internal connectivity, the
ensemble does not privilege any spatial direction and is therefore
statistically isotropic.

\paragraph{Diamond-like structures} were generated by perturbing the atomic positions of an
ideal diamond lattice within the unit cell.
Random displacements drawn from isotropic normal distributions were applied
uniformly to all lattice sites, with the displacement magnitude varied across
realizations.
Edges corresponding to nearest-neighbor bonds were then thickened to form struts
with randomly selected square cross sections, producing networks that remain
statistically isotropic at the ensemble level.

\paragraph{Cubic-strut structures} were generated using the same procedure as for the
diamond-like structures, but starting from a simple cubic lattice rather than a
diamond lattice.
Random perturbations of the lattice sites and variations in strut thickness
introduce geometric disorder while preserving the absence of any preferred
orientation in the ensemble.

\paragraph{Spline-based structures} were constructed by thresholding smooth, periodic scalar
fields defined via trivariate tensor-product B-splines.
Random spline coefficients were drawn independently, and periodicity was
enforced at the level of the spline basis.
Because the underlying scalar fields are generated without directional
preference, the resulting binary porous structures form a statistically
isotropic ensemble with smooth pore morphologies and controlled volume
fractions.

\subsection{Dataset of anisotropic transformed topologically diverse structures}
For the TD dataset, there is no natural or
straightforward way to introduce anisotropy at the level of structure
generation.
To overcome this limitation and to enable a systematic study of anisotropic
effects, a simple geometric elongation procedure was applied to the existing
binary structures, yielding an additional dataset denoted Anisotropic Transformed Topologically Diverse (ATTD).

Each original isotropic structure, represented on an $80\times80\times80$ voxel grid, was first downsampled to a $40\times40\times40$ grid by partitioning it into non-overlapping $2\times2\times2$ blocks and setting each coarse voxel to solid if any of its eight constituent voxels was solid (minority voting to guarantee continuity), forming a coarse-grained structural segment.
This segment was then elongated along the $z$ direction to obtain a
$40\times40\times80$ volume by duplicating each of its 40 layers along $z$ into two consecutive layers.

This transformation preserves the local topology of the original structures
while introducing a distinguished spatial direction through geometric
elongation.
As a result, isotropy is systematically broken and a moderate, controllable
structural anisotropy is introduced, with the $z$ axis becoming preferentially
aligned.
The resulting dataset provides a complementary benchmark bridging the gap
between statistically isotropic diverse structures and the strongly anisotropic
RTP-based datasets.

\subsection{Estimation of the elastic stiffness tensor}              \label{sec:fftmad}

Effective elastic properties of the porous microstructures were computed in silico using an FFT-based
homogenization approach following the augmented-Lagrangian scheme of Michel, Moulinec and
Suquet~\cite{Michel2001} (implementation available in the accompanying code repository; see "Code and
Data Availability"). Each voxelized structure is treated as a periodic solid/void unit cell, and for a
prescribed macroscopic strain the compatible microscopic strain field is found by iterating between a
local, voxel-wise elastic update and a projection onto compatible strains, the latter evaluated in
Fourier space via the Green's operator of a homogeneous reference medium. The void phase is modeled as a
highly compliant elastic medium (stiffness contrast 100 relative to the solid, chosen from a
convergence study) rather than as a true zero-stiffness phase, which regularizes the iteration; results
were cross-checked against the independent spectral solver \texttt{FFTMAD}~\cite{Lucarini2019} on a subset of structures. For each structure, the full effective stiffness tensor was obtained from six independent unit-strain load cases (three uniaxial and three pure-shear).

Under the assumption of linear elasticity and small strains, the relationship between the macroscopic stress and strain of the homogenized medium is given by the generalized Hooke's law 
$\sigma_i = \sum_{j=1}^{6} C_{ij}\,\varepsilon_j, \qquad i=1,\dots,6,$
where $\sigma_i$ and $\varepsilon_j$ are the stress and (engineering) strain components in Voigt notation and $C_{ij}=C_{ji}$ is the symmetric $6\times6$ effective stiffness tensor~\cite{Nye1957}. For a general (triclinic) anisotropic material, the tensor is fully characterized by 21 independent components $C_{ij}$, $i\le j$, with no material symmetry (e.g., orthotropy) imposed a priori.

To relate predictive performance to the physical role played by each elastic constant, the 21 independent components of $C_{ij}$ were grouped into five, partially overlapping subsets, used throughout the Results section to report and discuss prediction accuracy:

\begin{itemize}
\item \textbf{Uniaxial} ($C_{11}, C_{22}, C_{33}$): the three diagonal normal stiffnesses, relating a normal stress to the strain along
the same axis. These are the three-dimensional analogues of the uniaxial Young's modulus, and are the components most strongly and directly
governed by porosity and pore alignment along the loading direction.
\item \textbf{Poisson} ($C_{12}, C_{13}, C_{23}$): the off-diagonal normal--normal stiffnesses, coupling normal strain along one axis to
normal stress along an orthogonal axis. These terms govern the lateral expansion/contraction response of the material (Poisson-type
coupling) and depend on the correlation between pore morphology along different, mutually orthogonal directions.
\item \textbf{Shear} ($C_{44}, C_{55}, C_{66}$): the three diagonal shear stiffnesses, relating shear stress to shear strain in the $yz$,
$xz$, and $xy$ planes, respectively. They quantify the resistance of the porous network to pure shear deformation and are governed primarily
by the connectivity and bending resistance of the solid skeleton rather than by porosity alone.
\item \textbf{Off-diagonal} ($C_{14}, C_{15}, C_{16}, C_{24}, C_{25}, C_{26}, C_{34}, C_{35}, C_{36}, C_{45}, C_{46}, C_{56}$): the twelve
normal--shear and shear--shear coupling terms -- $C_{14}$ through $C_{36}$ couple a normal strain to a shear stress (or vice versa),
while $C_{45}, C_{46}, C_{56}$ couple two distinct shear planes to one another. 
Non-zero values of these components indicate coupling between normal and shear deformation, or between different shear modes, arising from structural asymmetry and/or misalignment of the coordinate axes with the material symmetry axes. In the datasets considered here, these components are generally smaller in magnitude than the principal stiffness components and prove substantially more difficult to predict.
\item \textbf{Full-tensor}: the union of all 21 independent components above, providing a complete characterization of the effective
elastic anisotropy of the structure.
\end{itemize}

The solid phase of the RTP structures was modeled as an isotropic linear elastic material with Young's modulus
$E_{\mathrm{bulk}}=81.5\,\mathrm{GPa}$ and Poisson's ratio $\nu_{\mathrm{bulk}}=0.39$, representative of the Au$_{0.30}$Ag$_{0.70}$
alloy~\cite{ShanShi, ZANDERSONS2021116979, BEETS2021116445}, while the TD and ATTD structures used $E_{\mathrm{bulk}}=70.0\,\mathrm{GPa}$
and $\nu_{\mathrm{bulk}}=0.33$, corresponding to aluminium. All simulations were performed in the small-strain regime on the voxelized
structures.

The resulting distributions of all 21 independent stiffness-tensor components across the three datasets
are compared in Section~S7 of the Supplementary Information.

\section{Computing topological descriptors on the structures}
\begin{figure}
    \centering
    \includegraphics[width=\textwidth]{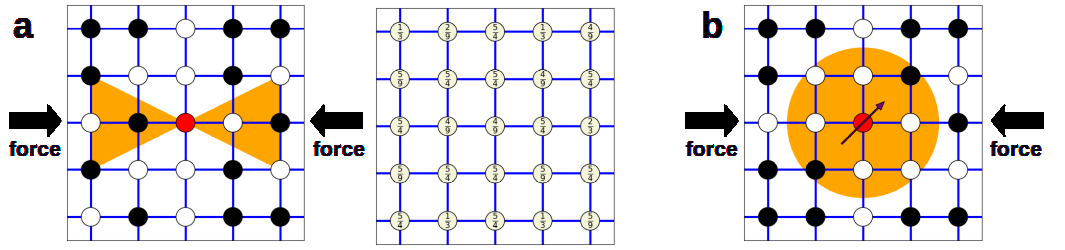}
    \caption{Diagrams \textbf{a} and \textbf{b} illustrate the calculation of filtration based on cones and principal components, respectively. The actual filtrations were calculated for three-dimensional grids, but for clarity, the examples are two-dimensional. Black and red circles represent non-empty fragments of material, while white circles represent empty spaces. In both cases, the filtration value is calculated for the central element marked in red.
    In the first image of diagram \textbf{a}, there are 5 non-empty and 4 empty elements in the constructed cones (orange triangles), so the filtration for the middle point is equal to $4/9$.
    By applying this procedure to other grid points, we obtain the grid of numbers presented in the second image in diagram \textbf{a}.
    In diagram \textbf{b}, if we limit ourselves only to elements contained in a certain neighborhood (orange circle), all non-empty elements will lie along one direction (marked with a purple arrow). This direction is the basis for calculating the multifiltration value.}
    \label{fig:TDA_directional}
\end{figure}

\subsection{Topological data analysis}

Topological data analysis (TDA) leverages the tools of topology and algebra to extract relevant information from datasets. The fundamental principle of this methodology is that "Data has shape and shape has meaning"~\cite{Dlotko2024}. The reader is referred to~\cite{Dlotko2024, Gurnari2025} for a comprehensive overview of the basic concepts. Here, we present a brief outline of the TDA methods we will use.

To represent the topology of the structures, we used 3-dimensional cubical complexes, which consist of cubes of various dimensions. TDA enables the analysis of data structures at multiple resolutions, which are defined using a function called filtration~\cite{Dlotko2024, Gurnari2025}. The initial values of this function were computed for 0-dimensional cubes (vertices) and then propagated to higher-dimensional cells in a standard manner, known as V-construction~\cite{Robins2011}.

The filtrations chosen in previous applications of TDA to porous materials are based on the distance transform and therefore do not account for directionality~\cite{Robins2016, Herring2019, Moon2019, Thompson2023, Lisitsa2020, Lee2017, Krishnapriyan2021, Chen2025, Wang2025, Jiang2018, Ishihara2023}. However, directionality is a critical factor for anisotropic materials whose Young's modulus depends on the direction of compression. To address this, we encoded the compression direction information into the filtration functions, thereby making our descriptors direction-aware. In addition to these direction-aware descriptors, we also compute non-directional (\emph{undirectional}) topological descriptors from a rotation-invariant Euclidean distance-transform filtration~\cite{Robins2016, Herring2019, Moon2019, Thompson2023, Lisitsa2020, Lee2017, Krishnapriyan2021, Chen2025, Wang2025, Jiang2018, Ishihara2023}: each voxel is assigned a signed Euclidean distance to the material/void interface, positive inside the solid phase and negative in the void, computed under the same periodic boundary conditions as the cone-based filtration; by construction this filtration does not depend on any loading direction. 
Both the direction-aware and the undirectional descriptors are evaluated throughout this study, which allows the specific contribution of embedding directionality into the filtration to be isolated. The details of these functions are described in the next section.

\subsubsection{Cone-based filtration}

The first directional filtration we consider is based on porosity inside cones. For each element of the input grid two cones starting at this point and having rotational axis parallel to direction of compressions are constructed. The cones are positioned in opposite directions. In the case of extreme grid elements, the cones may extend beyond the boundaries of the defined material. Therefore, for calculation purposes, we assume that the material is periodic. Porosity inside the cones is filtration value assigned to vertex corresponding to selected element of the grid. If the grid element represents an empty space, the corresponding filtration value is set to 1.25. The value 1.25 is selected arbitrary so that all the grid points outside material enters the filtration after all the grid elements belonging to material (that have filtration values between 0 and 1). This is to make a clear distinction between empty and non-empty spaces in the structure. The method is presented graphically in the Figure~\ref{fig:TDA_directional}. The formal definition of the described filtration is as follows.

\begin{definition}
A double cone with height $h$ and radius $r$ attached at point $e=(i_0, j_0, k_0)$ is defined as the following set of indices
\[ \mathrm{Cone}(e, r, h) = \left\{ (i, j, k) \in \mathcal{N}: |k - k_0| < h \land \sqrt{|i -i_0|^2 + |j -j_0|^2} \leq \frac{|k-k_0|r}{h}  \right\}  \]
\end{definition}

\begin{definition}
Let $G$ be a three-dimensional lattice (grid) of size $n \times n \times n$. The cone-based filtration associated with the vertex corresponding to the grid element $g_e$ is given by the expression
    \[F_{cone}(e; r,h) =1 - \frac{1}{|\mathrm{Cone}(e, r, h)|} {\sum_{(i,j,k) \in {\mathrm{Cone}}(e, r, h)}{g_{i'j'k'}}}, \]
    where $i' = i \mod n$, $ j' = j \mod n$, $ k' = k \mod n$ and $|\mathrm{Cone}(e, r, h)|$ is the number of voxels within the cone.
\end{definition}

Cone-based filtration distinguishes the direction of compression ($z$-axis here) and is invariant under rotation around an axis parallel to that direction. This method takes two parameters, the cone's height and radius of its base. To evaluate the filtration for an arbitrary loading direction $\hat d$, the grid is first re-oriented so that $\hat d$ coincides with the $z$-axis, and the cone-based filtration above is then computed on the re-oriented grid. For the three axis-aligned directions this re-orientation is an exact permutation of the grid axes; for the remaining, diagonal directions it is carried out by a rigid rotation of the binary grid with nearest-neighbour assignment (no interpolation).
In this work, the filtrations were evaluated independently for seven directions spanning the axis-aligned, face-diagonal, and body-diagonal directions of the cubic grid, written $[hkl]$ for the (unnormalized) direction vector $(h,k,l)$:
$
\hat d \in \big\{ [100],\,[010],\,[001],\,[110],\,[101],\,[011],\,[111] \big\},
$
namely the three Cartesian axes ($[100]$, $[010]$, $[001]$), the three face diagonals ($[110]$, $[101]$, $[011]$), and the body diagonal ($[111]$), each giving rise to an independent set of directional descriptors. For a fair comparison, all non-topological baseline descriptors were computed using the same set of seven directions (see Section S3 of the Supplementary Information).

\subsubsection{Filtration's extension based on principal component}
In the next step, we extended our initial filtration method to include information about the local material direction within specific regions. The local material direction for each material was represented by two parameters, which we determined through the following procedure.

For a selected grid element e, we treated all non-zero grid elements within a centered sphere of a radius $r$ as a 3D point cloud. We then calculated the first principal component  of this cloud to define its primary directional axis. The resulting normalized vector was used to extract two components: one corresponding to the compression direction ($v_z$) and a second, independent component ($v_y$). 

This process results in a 2-parameter filtration value, specifically $(1-|v_z|,1-|v_y|)$. 
This filtration is motivated by the principle that a material's ability to transfer stress is maximized when its orientation aligns with the imposed compression direction. The method requires a single hyperparameter: the radius of the point's neighborhood. As for the cone-based filtration, the compression direction $\hat d$ is embedded by first re-orienting the grid so that $\hat d$ coincides with the $z$-axis, and evaluating $v_z$ and $v_y$ on the re-oriented grid. A recap of this method is shown in the Figure {\ref{fig:TDA_directional}}

The hyperparameters used in the filtration procedures were fixed across all experiments. They were optimized on the Topologically Diverse (TD) dataset, the most versatile of the datasets considered in this study, and the resulting values were then applied unchanged to all datasets (see Section S4 of the Supplementary Information for the full hyperparameter scan). This experiment showed only a modest dependence of predictive performance on the specific hyperparameter values, indicating a stable picture across the ranges tested. For the cone-based filtration, the cone height was set to $h_{\mathrm{cone}} = 6$ and the cone base radius to $r_{\mathrm{cone}} = 3$. The same geometric parameters were employed for both the single-parameter filtration used in persistent homology and the multifiltration used in Euler characteristic profile calculations. In the principal-component-based multifiltration, the neighborhood radius was set to $r_{\mathrm{PC1}} = 3$, defining the spherical region used to estimate the local material orientation.

\begin{algorithm}
\caption{An algorithm of computing principal component-based multifiltration}\label{alg:cap}
\textbf{Input} \\
  \hspace*{\algorithmicindent}  $G$ - binary grid representing material \\
  \hspace*{\algorithmicindent}  $(i, j, k)$ - coordinates of selected element \\
  \hspace*{\algorithmicindent}  $r$ - neighborhood's radius
\begin{algorithmic}
\Procedure{PcFiltration}{$G,(i,j,k),r$}
\If{$G[i,j,k] = 0$}
    \State \Return (1.25, 1.25)
\Else
    \State $neighborhood \gets \mathrm{get\_neighborhood(G, (i,j,k), r)}$
    \State $v \gets \mathrm{calculate\_first\_principal\_component}(neighborhood)$
    \State $v \gets v / |v|$ \Comment{Normalization of 1st principal component}
    \State \Return $(1 - |v[2]|, 1 - |v[1]|)$
\EndIf
\EndProcedure
\end{algorithmic}
\end{algorithm}

\subsubsection{Direction-aware topological descriptors}

The filtrations described above were applied to study the structures at various resolutions. To generate topological summaries, we applied persistent homology to a cone-based filtration and Euler characteristic profiles (ECPs) to a three-parameter multifiltration combining cone- and principal component-based approaches. Persistent Homology (PH) is a fundamental tool in TDA. It offers the advantage of simultaneously capturing structural features, such as connected components, rings, and voids, at various scales. While ECP offers a less detailed topological description, it is computationally very efficient and allows for the use of multiparameter filtrations.

In the case of Persistent Homology calculations, the assumption of periodicity of the material in the  compression direction was used. Euler characteristic profiles were calculated without assuming periodicity in any direction.

Since most machine learning models require vector-formatted inputs, persistence diagrams were converted into persistence images~\cite{JMLR:v18:16-337}. Each diagram was mapped to a $12 \times 12$ image with birth and persistence ranges both set to $(0, 1.25)$. Persistence diagrams in dimensions 0, 1, and 2 were vectorized separately and subsequently concatenated into a single descriptor. Euler characteristic profiles were vectorized independently by sampling the multifiltration values on a $9 \times 9 \times 9$ grid 
The resulting PH and ECP descriptors are concatenated into a single combined descriptor (PH+ECP); both its direction-aware and undirectional variants, denoted "TDA" and "TDA (undir.)" throughout, are evaluated in Section~\ref{sec:results}.

\subsection{Estimation using boosting algorithm and neural networks}

Vectorized topological descriptors were used to train machine learning models aimed at predicting the elastic properties of the corresponding porous structures. Predictive models based on topological descriptors were trained using the CatBoost algorithm, a gradient-boosted decision tree method designed to provide strong performance and robustness on structured tabular data~\cite{Prokhorenkova2019}. As a baseline, we employed a convolutional neural network with the DenseNet-121 architecture, trained directly on voxelized representations of the structures. This architecture has previously been shown to achieve state-of-the-art performance for predicting mechanical properties of RTP-type porous structures~\cite{praca_z_madrytem}. 
\new{The network was trained from scratch, with the convolutional weights initialized from the He (Kaiming) normal distribution~\cite{He2015}.}
\new{The hyperparameters used to train both model families, and the sensitivity of the reported metrics to them, are given in the Supplementary Information, Section~S9.}

Model training was performed using $k$-fold cross-validation with $k=5$. The dataset was partitioned into five folds; for each of the five models, one fold (20\% of the data) was held out as the test set, while the remaining four folds were further split at random into a training set and a validation set in an 84:16 ratio (approximately 67\% and 13\% of the full dataset, respectively), the latter used to monitor overfitting and determine when to stop training. All metrics reported in this paper were computed exclusively on the test sets, and each of the five folds served as the test set exactly once. For each algorithm, a total of five models were trained, and the overall performance was obtained by averaging the metrics across all test folds. Using cross-validation increases the robustness of the results to random data splits and ensures that each data point is used exactly once as a test instance, allowing the reported metrics to represent the entire dataset.

Each structure contributes a single row containing all directional descriptors, for every direction considered, together with all independent components of the stiffness tensor, as columns. Consequently, a given structure's descriptors and elastic tensor components are always assigned jointly to the same cross-validation fold, and data leakage across directions or tensor components is avoided by construction.

\section{Results and Discussion}
\label{sec:results}

\begin{figure}
    \centering
    \includegraphics[width=\linewidth]{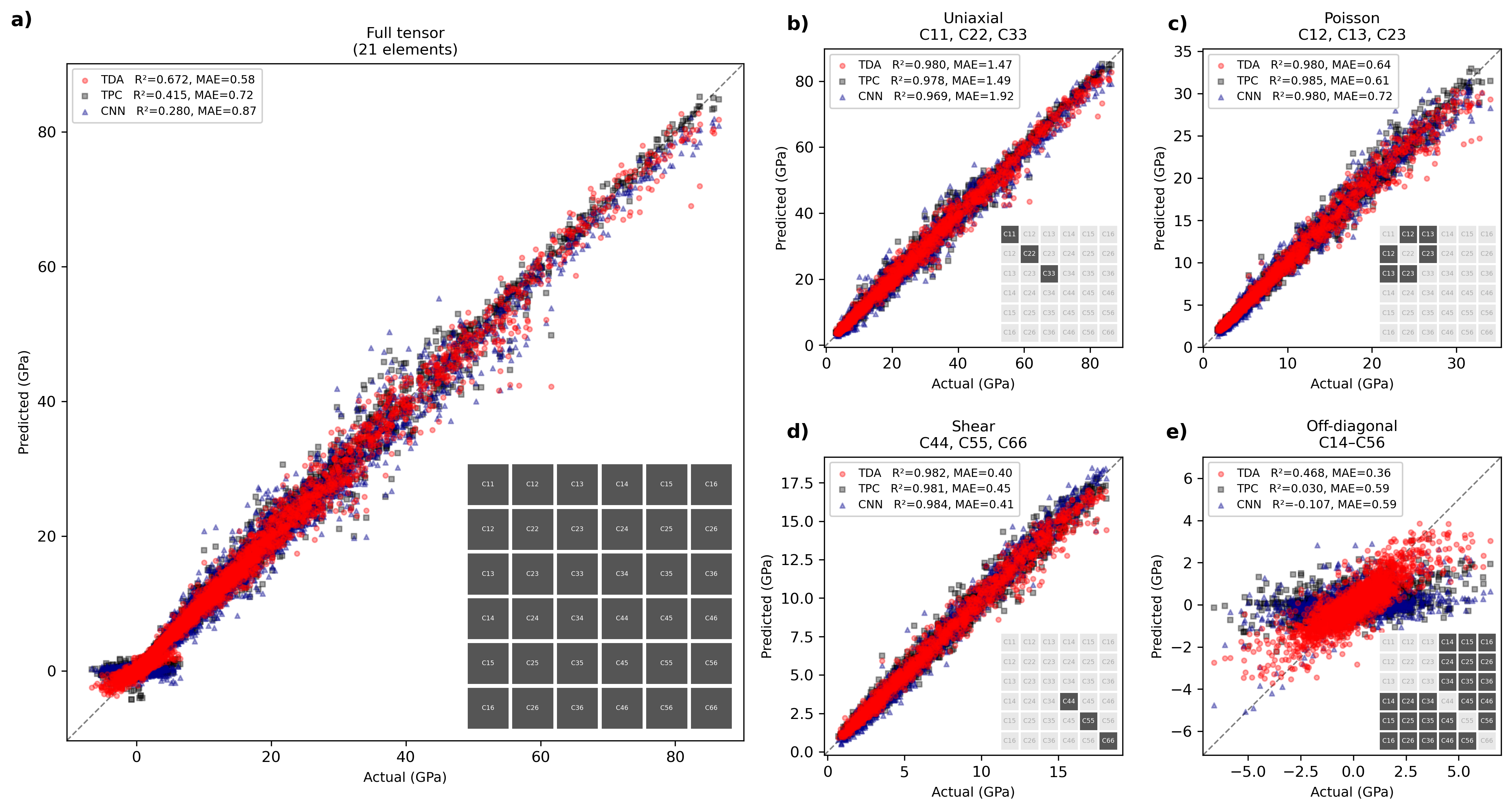}
    \caption{
    Predicted vs. actual stiffness-tensor components on the RTP dataset, comparing TDA (red circles), TPC (black squares), and CNN (dark-blue triangles). Each panel pools test-set predictions from all 5 cross-validation folds; the dashed line marks perfect prediction. $R^2$ and MAE is averaged across the 5 CV folds. The inset in the lower-right corner of each panel is a schematic of the $6 \times 6$ Voigt stiffness matrix, with the independent elements belonging to that panel's target group shaded dark gray. a) Full tensor: all 21 independent elements pooled together. b) Uniaxial group: diagonal normal-stress terms $C_{11}$, $C_{22}$, $C_{33}$. c) Poisson group: off-diagonal normal-normal coupling terms $C_{12}$, $C_{13}$, $C_{23}$. d) Shear group: diagonal shear terms $C_{44}$, $C_{55}$, $C_{66}$. e) Off-diagonal group: the 12 normal-shear and shear-shear coupling terms. 
    \new{Note that the vertical scales differ between panels: the off-diagonal components in panel~(e) are one to two orders of magnitude smaller than the diagonal components in panels~(b--d), so a given absolute deviation corresponds to a much larger relative error there.}     
    }
    \label{fig:scatter_rtp}
\end{figure}

\begin{figure}
    \centering
    \includegraphics[width=\linewidth]{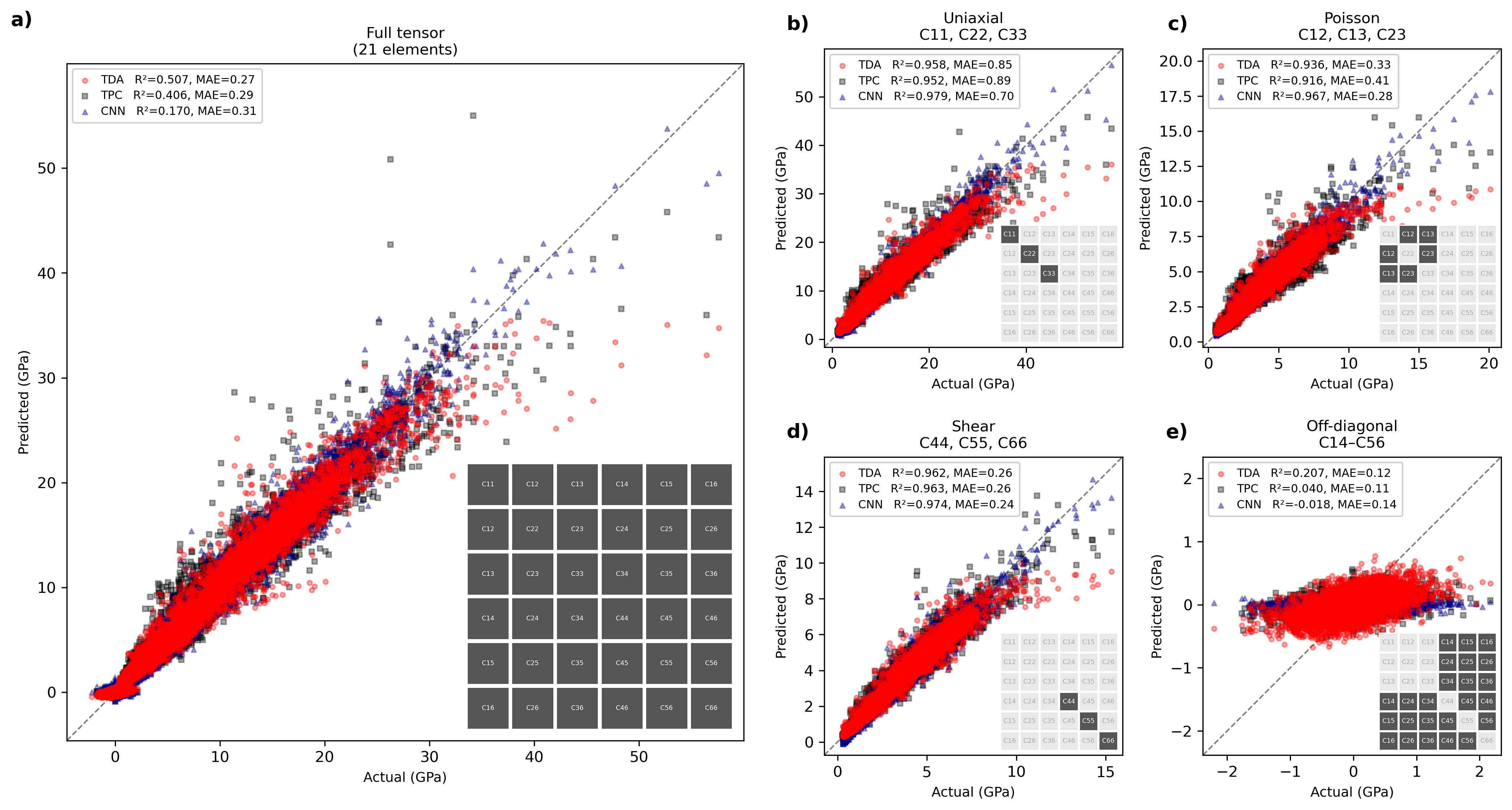}
    \caption{
    The same as Figure \ref{fig:scatter_rtp} but for the TD dataset.
    }
    \label{fig:scatter_td}
\end{figure}

Table~\ref{tab:metrics_rtp} reports the $R^2$ obtained on the RTP dataset for each of the five stiffness-tensor target groups defined above
(Uniaxial, Poisson, Shear, Off-diagonal, and Full-Tensor), for four baseline descriptors (Porosity, the two-point correlation function TPC,
the Fabric tensor, and a DenseNet-121 CNN operating directly on the voxelized structure) together with the direction-agnostic (TDA (undir.))
and direction-aware (TDA) topological descriptors introduced above. For Porosity, TPC, Fabric, and TDA, the descriptor set supplied to each
model comprises the corresponding features computed along all seven directions considered in this study (the three coordinate axes, the
three face diagonals, and the body diagonal of the unit cell), so that no directional information available from the sampled orientations is
withheld from any method; the CNN, by contrast, is trained directly on the full voxelized structure and therefore has access to the complete
geometry regardless of direction. For every method and every target group, an independent model was trained from scratch using only the
corresponding descriptor set (or, for the CNN, the voxel grid) as input and the matching subset of Voigt components as the regression
target. 
For target groups comprising more than one Voigt component (Uniaxial, Poisson, Shear, Off-diagonal, and Full-Tensor), the $R^2$ reported for a given cross-validation fold is the average of the per-component $R^2$ values, each computed independently on that fold's test set for the corresponding Voigt component. Reported values in Table~\ref{tab:metrics_rtp} are the mean and standard deviation of this per-fold, per-group $R^2$ across the five cross-validation folds used throughout this study.
Table~\ref{tab:metrics_td} reports the analogous results for the topologically diverse (TD) dataset, and Figures~\ref{fig:scatter_rtp} and
\ref{fig:scatter_td} show the corresponding predicted-versus-actual scatter plots, panelled by target group, for the three most competitive
methods (TDA, TPC, and CNN). Corresponding mean absolute error (MAE) values, a scale-dependent complement to $R^2$, are reported for
both datasets in Section S4 and S5 of the Supplementary Information (Tables~S1, S2 and S4).
\new{
Metrics of this kind depend to some extent on the training configuration; repeating the cross-validation under variations of the data partition, the learning rate, the number of boosting iterations and the tree depth changes $R^2$ by at most $0.012$ for the diagonal groups and by up to $0.09$ for the Off-diagonal group, and the CNN baseline, which does involve weight initialization, varies more but stays indistinguishable from zero on the Off-diagonal group under
every configuration tested (Supplementary Information, Section~S9). These variations are small compared with the differences between descriptor families discussed below and leave their ordering unchanged.
}

On the RTP dataset, the three "diagonal" target groups -- Uniaxial, Poisson, and Shear -- are predicted with comparably high accuracy by
Porosity, TPC, the CNN, and both TDA variants, with $R^2$ ranging between 0.94 and 0.98 (Table~\ref{tab:metrics_rtp}); only the Fabric
tensor performs markedly worse. Within this narrow band, directional TDA is nevertheless the
best-performing method on Uniaxial ($R^2=0.980$, versus $0.978$ for TPC) and the second-best on Shear ($R^2=0.982$, versus $0.984$ for the
CNN). This close agreement is visible directly in Figure~\ref{fig:scatter_rtp}(b--d), where the point clouds for TDA, TPC, and CNN all
collapse tightly onto the identity line with no visually distinguishable difference between methods. Even where all competitive methods are
closely matched, direction-aware topology is therefore never worse than second.

The picture changes significantly for the Off-diagonal group, comprising the twelve normal-shear coupling terms. Here, directional TDA achieves $R^2=0.468$, far ahead of the second-best method, the Fabric tensor ($R^2=0.287$), while
Porosity, TPC, and the CNN all remain close to zero or negative ($R^2=-0.013$, $0.030$, and $-0.107$, respectively).
Figure~\ref{fig:scatter_rtp}(e) makes this gap immediately apparent: the TDA predictions (red) retain a clear, positive
correlation with the actual values, whereas the TPC (grey) and CNN (blue) predictions collapse into a nearly horizontal band clustered
around zero, essentially uninformative about the sign or magnitude of the true coupling term. Because the Full-Tensor group pools all 21
independent components -- the nine well-predicted diagonal terms together with the twelve off-diagonal ones -- the large advantage of
directional TDA on the off-diagonal terms propagates directly into the Full-Tensor column, where it again achieves the best overall accuracy
($R^2=0.672$), roughly 60\% higher (in relative terms) than the second-best method, TPC ($R^2=0.415$). This is consistent with
Figure~\ref{fig:scatter_rtp}(a): because the pooled Full-Tensor panel is dominated by the nine well-predicted diagonal components, the point
clouds for TDA, TPC, and CNN overlap almost completely across the full stiffness range, so the sizeable gap in $R^2$ between methods is not
visually apparent in this panel; it only becomes evident once the off-diagonal components are isolated in panel (e). The Off-diagonal group
is thus the primary driver of the overall advantage of direction-aware topology, and the Full-Tensor metric is best understood as inheriting
this advantage rather than reflecting an independent effect.
\new{The two blocks also differ in how absolute and relative error relate: the twelve off-diagonal
components have mean magnitudes of $0.18$--$1.28$~GPa, that is, $1$--$3\%$ of $C_{11} \approx 46$~GPa, and are distributed around zero, so their absolute errors remain small (mean absolute
error $0.17$--$0.65$~GPa).}

The comparison between directional and undirectional TDA descriptors reinforces this picture. On the three diagonal target groups, removing
directional information from the filtration produces only a moderate reduction in accuracy (e.g., Uniaxial: $0.980 \to 0.943$). On the
Off-diagonal group, however, the undirectional descriptor collapses to $R^2=-0.018$, statistically indistinguishable from the weakest
baselines, and this collapse is again inherited by the Full-Tensor group ($0.672 \to 0.341$). The benefit of embedding the loading direction
into the filtration is therefore not distributed uniformly across the elastic tensor, but is concentrated precisely in the coupling terms
that a rotation-invariant, direction-agnostic filtration cannot in principle resolve.

The same qualitative pattern is reproduced on the TD dataset (Table~\ref{tab:metrics_td}, Figure~\ref{fig:scatter_td}), which -- unlike RTP
-- comprises five topologically distinct structure families (Voronoi, Zeolitic, Diamond-like, Cubic-strut, and Spline-based) generated
independently of any preferred spatial direction, and is therefore substantially more diverse in pore morphology and connectivity. On the
diagonal target groups, the CNN is now the strongest method (Uniaxial $R^2=0.979$, Poisson $0.967$, Shear $0.974$), with directional TDA a
close second on Uniaxial and Poisson ($0.958$ and $0.936$) and effectively tied for second on Shear ($0.962$, against $0.963$ for TPC);
Figure~\ref{fig:scatter_td}(b--d) shows that all three methods again lie close to the identity line, though the CNN and TPC scatter plots
reveal a handful of large overpredictions at the highest stiffness values that are largely absent from the TDA predictions. On the
Off-diagonal group, however, directional TDA again dominates by a wide margin ($R^2=0.207$, against $0.040$ for the next-best method, TPC),
a margin that carries over to the Full-Tensor group ($0.507$, against $0.406$ for TPC); as in the RTP case, Figure~\ref{fig:scatter_td}(e)
shows TDA retaining a visible, if weak, trend along the identity line while TPC and CNN again flatten into an uninformative band near zero.
As on RTP, the undirectional descriptor trails the directional one throughout, with the largest gap again occurring on Off-diagonal
($-0.003$ versus $0.207$) and Full-Tensor ($0.351$ versus $0.507$). These results confirm that the advantage of direction-aware topology on
the coupling terms of the elastic tensor is not specific to the RTP structure-generation process, but generalizes to a substantially more
diverse set of porous topologies.

As a further robustness check, we additionally evaluated the same set of methods and target groups on an auxiliary Anisotropic Transformed
Topologically Diverse (ATTD) dataset, obtained by elongating the TD structures along a single axis to introduce moderate, controlled
anisotropy while preserving their topological diversity. The corresponding $R^2$ and MAE summary tables, together with the
predicted-versus-actual scatter plots analogous to Figures~\ref{fig:scatter_rtp} and \ref{fig:scatter_td}, are reported in Section S5 of the
Supplementary Information (Tables~S3, S4 and Figure S3).

Taken together, the two datasets place directional TDA among the top two methods in four of the five target groups. On RTP it is the best
method in three groups (Uniaxial, Off-diagonal, Full-Tensor) and second-best in one (Shear); on TD it is the best in two groups
(Off-diagonal, Full-Tensor) and second-best in two more (Uniaxial, Poisson). The only method that outperforms directional TDA more broadly
is the CNN on the TD dataset, which wins the three diagonal target groups outright. This is noteworthy in itself: the CNN is not a
feature-based model -- it is trained end-to-end directly on the raw voxel grid, without any hand-engineered structural descriptor, and
therefore has access, in principle, to strictly more information than any of the descriptor-based methods. Yet on both datasets the CNN
fails on precisely the target group where directional TDA excels, remaining close to zero or negative on Off-diagonal ($R^2=-0.107$ on RTP,
$-0.018$ on TD) and consequently weak on Full-Tensor ($0.280$ and $0.170$), as seen directly in panel (e) of Figures~\ref{fig:scatter_rtp}
and \ref{fig:scatter_td}. That a compact, physically interpretable, direction-aware topological descriptor matches or exceeds an
unconstrained end-to-end deep model on the hardest-to-predict components of the elastic tensor, across two structurally unrelated datasets,
is a central finding of this study. 
Finally, to assess whether the $R^2$ differences reported above reflect genuine, statistically robust
effects rather than incidental variation across cross-validation folds, we performed one-sided Wilcoxon
signed-rank tests comparing, structure by structure, the absolute prediction error of directional TDA
against every baseline method, separately for each dataset and target group (Section~S6 of the
Supplementary Information). These tests confirm the picture above at the level of individual structures:
directional TDA is significantly more accurate than every non-CNN baseline across all datasets and target
groups, and significantly more accurate than the CNN specifically on the Off-diagonal (and, for TD,
Full-Tensor) groups; conversely, the CNN's edge on the diagonal groups is likewise statistically
significant. The observed differences are therefore systematic rather than artifacts of a particular
cross-validation split.

\begin{table}[]
\centering
\caption{Prediction accuracy ($R^2$, mean(std) across the 5 CV folds) for each stiffness-tensor target group on the RTP dataset. Three descriptor-based baselines (Porosity, TPC, Fabric) and a CNN operating directly on the voxelized structure against the direction-agnostic TDA (undir.) descriptor and our direction-aware TDA descriptor. Target groups follow the same grouping as Figure~\ref{fig:scatter_rtp}. For each column, the best result is shown in \textbf{bold} and the second-best in \textit{italic}.}
\label{tab:metrics_rtp}
\begin{tabular}{l|rrrrr}
\multicolumn{1}{l}{Method} & \multicolumn{1}{c}{Uniaxial} & \multicolumn{1}{c}{Poisson} & \multicolumn{1}{c}{Shear} & \multicolumn{1}{c}{Off-diagonal} & \multicolumn{1}{c}{Full-Tensor} \\ \hline \hline
Porosity & 0.969(.009) & \textit{0.982(.005)} & 0.974(.008) & -0.013(.010) & 0.393(.014) \\
TPC & \textit{0.978(.008)} & \textbf{0.985(.005)} & 0.981(.006) & 0.030(.049) & \textit{0.415(.024)} \\
Fabric & 0.018(.036) & 0.021(.043) & 0.031(.043) & \textit{0.287(.021)} & 0.119(.046) \\
CNN & 0.969(.013) & 0.980(.007) & \textbf{0.984(.005)} & -0.107(.212) & 0.280(.112) \\ \hline
TDA (undir.) & 0.943(.016) & 0.955(.011) & 0.949(.014) & -0.018(.019) & 0.341(.019) \\
TDA & \textbf{0.980(.007)} & 0.980(.007) & \textit{0.982(.006)} & \textbf{0.468(.019)} & \textbf{0.672(.016)}
\end{tabular}
\end{table}

\begin{table}[]
\centering
\caption{The same as Table \ref{tab:metrics_rtp} but for the TD dataset.}
\label{tab:metrics_td}
\begin{tabular}{l|rrrrr}
\multicolumn{1}{l}{Method} & \multicolumn{1}{c}{Uniaxial} & \multicolumn{1}{c}{Poisson} & \multicolumn{1}{c}{Shear} & \multicolumn{1}{c}{Off-diagonal} & \multicolumn{1}{c}{Full-Tensor} \\ \hline \hline
Porosity & 0.906(.003) & 0.892(.007) & 0.954(.004) & -0.008(.010) & 0.384(.007) \\
TPC & 0.952(.007) & 0.916(.019) & \textit{0.963(.009)} & \textit{0.040(.026)} & \textit{0.406(.003)} \\
Fabric & 0.452(.034) & 0.392(.032) & 0.444(.033) & 0.023(.010) & 0.197(.015) \\
CNN & \textbf{0.979(.007)} & \textbf{0.967(.008)} & \textbf{0.974(.014)} & -0.018(.018) & 0.170(.167) \\ \hline
TDA (undir.) & 0.848(.015) & 0.869(.017) & 0.910(.010) & -0.003(.001) & 0.351(.017) \\
TDA & \textit{0.958(.006)} & \textit{0.936(.010)} & 0.962(.004) & \textbf{0.207(.014)} & \textbf{0.507(.005)}
\end{tabular}
\end{table}

\section{Summary}
\label{sec:summary}

Direction-dependent properties in porous materials pose a structural learning problem in which symmetry breaking with respect to the loading axis is essential. This work demonstrates that incorporating directional information into topological data analysis provides a significant advantage for predicting direction-dependent elastic properties of porous structures, from a single directional Young's modulus to the full anisotropic stiffness tensor. Standard topological descriptors, while effective at capturing multiscale connectivity and void structure, are inherently isotropic and therefore insufficient for modeling anisotropic mechanical response. By embedding the loading direction directly into the filtration functions used for persistent homology and Euler characteristic profiles, the proposed direction-aware construction lifts this intrinsic isotropy and encodes mechanically relevant anisotropic topology in a compact representation.

Extending the analysis beyond a single scalar modulus to the full elastic stiffness tensor reveals that this advantage is highly
structured rather than uniform. On the diagonal groups of the tensor -- the uniaxial, Poisson, and shear terms -- direction-aware
descriptors, non-directional baselines, and a voxel-based CNN all perform comparably well; the decisive difference emerges on the
twelve off-diagonal, normal-shear coupling terms, where only direction-aware topology retains meaningful predictive power, while the
non-directional descriptors and the CNN collapse toward chance-level accuracy. This pattern, and the resulting advantage on the
pooled full-tensor metric, is reproduced across three datasets and is confirmed by paired, structure-by-structure Wilcoxon signed-rank tests rather than aggregate averages alone.
Notably, this directional advantage is not confined to strongly anisotropic structures: even on the topologically diverse (TD)
dataset, whose structures are generated independently of any preferred spatial direction, removing directional information from the
descriptor collapses its predictive power specifically on the off-diagonal terms, confirming that direction-aware topology captures
mechanically relevant organization that persists even in nominally isotropic microstructures. The direction-aware descriptors match
or exceed the CNN specifically on these hardest-to-predict coupling terms, despite the CNN having access, in principle, to
substantially more information via the raw voxel grid.

From the perspective of porous-materials modeling, these results establish a general, geometry-agnostic route for encoding anisotropic connectivity and alignment effects—key determinants of stiffness—without relying on handcrafted geometric descriptors or expensive end-to-end training on voxel grids. More broadly, the proposed direction-aware TDA framework provides a transferable and low-dimensional alternative (or complement) to voxel-based deep learning for structure-property prediction in porous media -- as demonstrated here for the full elastic stiffness tensor -- with natural extensions to other direction-dependent properties such as yield strength, permeability, and transport coefficients governed by anisotropic microstructural organization.

\new{
The framework is not intended to replace FFT-based homogenization, which is efficient, accurate, and is the method used here to generate the ground-truth tensors. Once trained, however, it predicts the full tensor about twenty times faster per structure, a single descriptor serving all $21$ independent components (Supplementary Information, Section~S8). This makes it useful for ranking large candidate populations -- in inverse design, or in shortlisting a database for subsequent verification with the solver.    
}

All conclusions in this study are drawn from computational experiments on synthetic, periodic, voxelized microstructures:
the anisotropic RTP dataset, in which anisotropy is controlled directly at the level of structure generation, and the
statistically isotropic TD dataset, spanning five geometrically unrelated pore-morphology families (an auxiliary
anisotropic dataset, ATTD, obtained by a geometric elongation transform, is discussed in the Supplementary Information).
Validating the proposed direction-aware descriptors against experimentally imaged or naturally anisotropic porous
microstructures is an important direction for future work.

\subsection*{Code and Data Availability}
A complete implementation of all methods used in this study is available in the accompanying GitHub repository at \url{https://github.com/dioscuri-tda/direction-aware-tda-for-porous-materials}. The repository includes the full codebase for computing directional and non-directional topological descriptors, the FFT-based solver used to compute the effective stiffness tensors (Section~\ref{sec:fftmad}), training the CatBoost and CNN models, reproducing all numerical experiments, and generating the figures reported in the manuscript. It also contains all datasets used in this work, including voxelized porous structures stored as \texttt{.npy} files, the computed effective stiffness-tensor values, and vectorized PH, ECP, and PH+ECP descriptors. Ready-to-use databases, training scripts, and detailed instructions for reproducing every result are provided to ensure full transparency and straightforward replication of the presented findings.

\subsection*{Funding}
Financial support from the PORMETALOMICS project, funded by the Spanish
Ministry for Science, Innovation, and Universities (award no.
PCI2022-132975) and the National Science Centre, Poland (project no.
2021/03/Y/ST5/00232) within the M-ERA.NET 3 call, is gratefully acknowledged.
This project has received funding from the European Union’s Horizon 2020 research and innovation programme under grant agreement No 958174.
P.D. was supported by the European Union's Horizon Europe research and innovation programme under the Marie Skłodowska-Curie (MSCA) grant agreement No. 101120290 (GAP).
Funded by the European Union. Views and opinions expressed are however those of the author(s) only and do not necessarily reflect those of the European Union or European Research Executive Agency. Neither the European Union nor the granting authority can be held responsible for them

This research was funded in whole or in part by the National Science Centre, Poland, grant number 2021/03/Y/ST5/00232. For the purpose of Open Access, the author has applied a CC-BY public copyright licence to any Author Accepted Manuscript (AAM) version arising from this submission.

This research is financed under Dioscuri, a programme initiated by the
Max Planck Society, jointly managed with the National Science Centre
in Poland, and mutually funded by Polish Ministry of Science and
Higher Education and German Federal Ministry of Research,
Technology and Space.

The calculations were made with the support of the Interdisciplinary Center for Mathematical and Computational Modeling of the University of Warsaw (ICM UW) under the computational grant no g105-2705.

\section*{Declarations}
\subsection*{Author Contributions Statement}
R.T., M.B., and J.M. wrote the main manuscript. R.T., M.B. generated the structures,  B.N. generated TD/spline structures. R.T. and M.B. performed the stiffness-tensor simulations. R.T. developed the machine-learning code and ran computations. 
M.B., J.M., R.T. and P.D. designed and implemented the topological data analysis framework. R.T., M.B., J.M. and P.D. contributed to the methodology. P.D., M.H., and B.N. provided computational resources. P.D. and M.H. supervised project administration. 
R.T. prepared all figures except Figure 3, made by J.M. and M.B. 
R.T. prepared the Supplementary Information except section S1
made by M.B. R.T. prepared the revised version of the manuscript and supplementary information. All authors contributed to formal analysis, validation of the results, and reviewed the final version of the manuscript.

\subsection*{Competing Interests}
The authors declare no competing interests.

\clearpage
\newgeometry{margin=1in}
\begin{center}
\Huge{Supplementary Information} 
\vspace{2cm} \\
\LARGE{Direction-aware topological descriptors for elastic stiffness tensor prediction in porous materials}
\end{center}

\begin{center}
\end{center}

\setcounter{section}{0}
\setcounter{figure}{0}
\setcounter{table}{0}

\renewcommand{\thesection}{S\arabic{section}}
\renewcommand{\thesubsection}{\thesection.\arabic{subsection}}
\renewcommand{\thefigure}{S\arabic{figure}}
\renewcommand{\thetable}{S\arabic{table}}

\etocsetlocaltop.toc{part}

\etocsetnexttocdepth{subsection}
\localtableofcontents

\newpage

\section{Topologically Diverse (TD) Dataset}
The RTP dataset is described in detail in the main text. All porous structures used in this study—including the RTP, TD, and ATTD datasets—are openly available in the dedicated repository associated with this paper: \url{https://github.com/dioscuri-tda/direction-aware-tda-for-porous-materials}. 
The repository also provides a utility for converting the voxelized data into VTK format, enabling straightforward visualization and rendering using standard tools such as ParaView~\cite{Ahrens2005ParaViewAE}.

\subsection{Generation of porous structures}

\subsubsection{Random Voronoi structures}\label{app:voronois}

\begin{figure}
    \centering
    \includegraphics[width=0.89\linewidth]{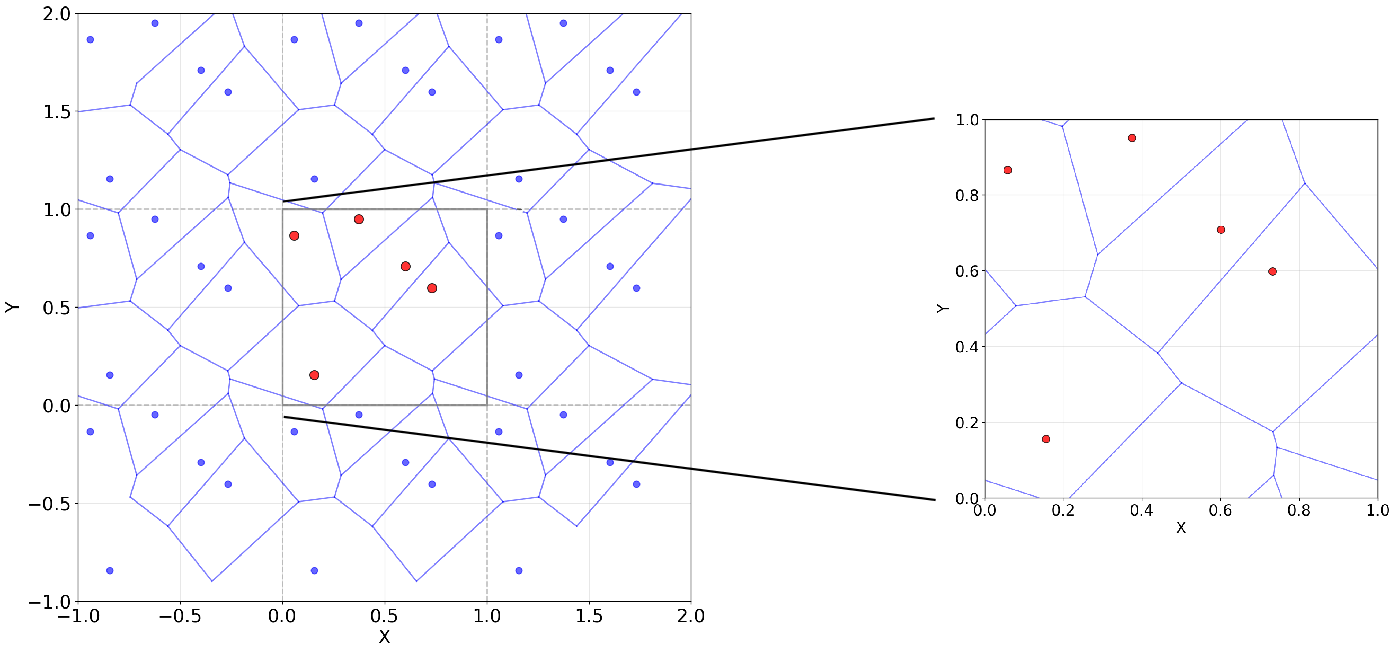}
    \caption{A 2D visualisation of the algorithm for generating random Voronoi structures satisfying periodic boundary conditions.}
    \label{fig:voronoi}
\end{figure}

600 random Voronoi structures were generated. For each structure, between 3 and 12 points were sampled from a unit cell (randomly, uniform distribution), with the space between them tesselated into Voronoi regions and the edges between regions used as a skeleton for the structure. The voxels traversed by the edges are approximated using the Bresenham's Line Algorithm~\cite{bresenham1965algorithm} and included within the structure. The edges are thickened to form struts with a square-shaped cross-section, with the side length of the square randomly drawn from the uniform distribution between 0.06 and 0.14. In order to guarantee periodicity of the structure, copies of the original points are generated in cells neighbouring the unit cell as well and the Voronoi tesselation of space, voxelisation of edges and subsequent thickening is conducted within the neighbouring cells as well (see Fig. \ref{fig:voronoi} for a visualization of a 2D version of the procedure), with the neighboring cells discarded at the last stage of the procedure. 

\subsubsection{Zeolitic-inspired structures}\label{sec:zeolites}

More than 3000 predicted zeolitic structures were recovered from Michael Deem's database~\cite{C0CP02255A, deem2023pcod}. Each structure is described by the basis cell vector and the locations of Si and O atoms in a unit cell. We restricted the studies to structures with a perpendicular shape of the unit cell, which still left over 1500 structures. We transformed the structures from perpendicular to the normalized $1\times1\times1$ cubic volume by stretching/compression. We thickened the edges by dilation to form struts with a square-shaped cross-section of side length 0.12. We used 300 randomly drawn structures from the dataset and connected the locations of the Si atoms in these structures after the stretching/compression to the locations of their nearest Si neighbors (with a cut-off distance of 4.0\AA), O atoms were not used in the structure. The connections formed the skeleton of the structure. Periodic boundary conditions were guaranteed by a procedure similar to the one used for random Voronoi structures, involving generating copies of the original Si atom locations in cells neighbouring the unit cell and connecting them to their nearest neighbours as well. Another 281 structures were created on the basis of different randomly drawn structures from using Michael Deem's database~\cite{C0CP02255A, deem2023pcod} using the same algorithm, but with using 0.04 instead of 0.12 as the side length of the square-shaped cross-section of the struts.  
This gives a total of 581 zeolite structures.

\subsubsection{Diamond structures}

299 diamond-resembling structures were generated. Each structure was generated by creating within the unit cell 8 vertices corresponding to atoms in a diamond lattice and creating between the vertices edges corresponding to bonds between the atoms. The precise locations of the vertices were determined by adding a stochastic displacement to the location of the corresponding atom in the diamond lattice. Each of the three-dimensional components of the stochastic displacement for all vertices in a given lattice were drawn from a normal distribution with a standard deviation $\epsilon$ itself drawn from a uniform distribution between 0 and 0.1 (compare to 1.0 being the length of the side of a unit cell). That way, the amount of stochasticity was varied between the structures. The edges connecting the vertices were thickened via dilation to form struts with square-shaped cross-sections, with the length of the side of the cross-section drawn from a uniform distribution between 0.06 and 0.26. That way, the volume fraction was varied between structures.

\subsubsection{Cubic structures}

300 structures resembling a simple cubic strut were generated using the procedure used for diamond structures, with the only difference the 8 initiating points are not the locations of atoms in a diamond structures like in the former dataset, but instead the following set of points: [0, 0, 0],
        [0.5, 0, 0],
        [0, 0.5, 0],
        [0, 0, 0.5],
        [0, 0.5, 0.5],
        [0.5, 0, 0.5],
        [0.5, 0.5, 0],
        [0.5, 0.5, 0.5]. Replicating further steps from the procedure used to generate the diamond-like structures leads to structures resembling a classical cubic strut, with varying thicknesses of the struts and levels of randomization in locations of the struts' vertices.

\subsubsection{Splines}

To generate a family of smooth porous morphologies, we construct a random \emph{periodic} scalar field on the unit cell and then threshold it.
Let $n=\texttt{sampling}$ be the number of spline control points per direction and draw i.i.d.\ coefficients 
\begin{equation}
r_{ijk}\sim\mathcal U(0,1),\qquad i,j,k=1,\dots,n .
\end{equation}
In \textsc{Mathematica}, the command
\texttt{BSplineFunction} builds a trivariate tensor-product B-spline field
(with \texttt{SplineDegree->3}, \texttt{SplineClosed->True}, \texttt{SplineKnots->"Unclamped"}), i.e.
\begin{equation}
f(x,y,z)=\sum_{i=1}^{n}\sum_{j=1}^{n}\sum_{k=1}^{n}
r_{ijk}\,N_{i,3}(x)\,N_{j,3}(y)\,N_{k,3}(z),
\label{eq:spline_field}
\end{equation}
where $\{N_{i,3}\}$ are cubic B-spline basis functions (defined via the Cox--de Boor recursion), and the ``closed'' setting enforces periodicity across the cell boundaries \cite{deBoor1978,wolframBSplineFunction}.

A binary structure is then obtained as a super-level set. Using wrapped (periodic) coordinates
$\tilde{{x}}=(x\bmod 1,\,y\bmod 1,\,z\bmod 1)$ (implemented with \texttt{Mod}), we voxelize on an $L\times L\times L$ grid (here $L=80$) by
\begin{equation}
T_{abc}(t)=\mathbf{1}\Bigl\{\,f\!\bigl(\tfrac{a}{L},\tfrac{b}{L},\tfrac{c}{L}\bigr)\ge t\,\Bigr\},
\qquad a,b,c=0,\dots,L-1,
\label{eq:thresholding}
\end{equation}
and report the volume fraction
\begin{equation}
\phi(t)=\frac{1}{L^{3}}\sum_{a,b,c} T_{abc}(t).
\end{equation}
Varying the threshold $t$ (in our implementation, on an approximately uniform grid in $[0.35,0.50]$) produces two independent subsets of spline-based structures each, spanning a range of $\phi(t)$ while preserving smoothness and periodic tiling of the unit cell.
Total of 596 spline structures were used in the dataset.

\section{TDA Hyperparameter Selection}
\label{sec:tda_hyperparams}

The direction-aware topological descriptors described in the main text depend on a small number of hyperparameters: the cone height $h_{\mathrm{cone}}$ and cone base radius $r_{\mathrm{cone}}$ of the cone-based filtration, the neighborhood radius $r_{\mathrm{PC1}}$ of the principal-component-based multifiltration, and the resolution used to vectorize each resulting topological summary into a fixed-length feature vector. Rather than tuning these hyperparameters separately for every dataset and every target, a single systematic scan was performed on the Topologically Diverse (TD) dataset, selected as the most versatile of the datasets considered in this study, and the resulting values were then carried over, unchanged, to the RTP and ATTD datasets.

\subsection{Parameters varied}

Two groups of filtration hyperparameters were scanned independently. First, the neighborhood radius of the principal-component-based multifiltration was varied as $r_{\mathrm{PC1}} \in \{2,3,4,6\}$, with the cone geometry held fixed at $h_{\mathrm{cone}}=6$, $r_{\mathrm{cone}}=3$. Second, the cone geometry itself was varied over three settings, $(h_{\mathrm{cone}}, r_{\mathrm{cone}}) \in \{(2,6),\,(4,4),\,(6,3)\}$, with the neighborhood radius held fixed at $r_{\mathrm{PC1}}=4$. In addition, two vectorization hyperparameters were scanned independently of the filtration parameters above: the grid resolution used to vectorize the Euler characteristic profile (ECP), $g \in \{4,6,8,10\}$, and the persistence-image resolution used to vectorize persistent homology (PH), $N \times N$ with $N \in \{6,8,10,12,16\}$.

For every combination of filtration and vectorization hyperparameters, a CatBoost model was trained on the TD dataset to predict the full stiffness tensor (\emph{full-tensor} target group) using all seven directions, following the same 5-fold cross-validation protocol described in Section~3 in the main text. The coefficient of determination ($R^2$), averaged over the five test folds, was used as the comparison metric throughout.

\subsection{What the vectorization grid resolution means}

Both ECP and PH ultimately require converting a filtration, defined pointwise on the voxel grid, into a fixed-length numerical descriptor. The two vectorization schemes discretize a different underlying space, and the role of their respective resolution parameters differs accordingly.

For the multifiltration used in ECP, every voxel carries three filtration values (the cone value and the two principal-component-derived directional components). The Euler characteristic profile summarizes the structure by evaluating the multiparameter Euler characteristic on a regular grid of threshold combinations spanning the range of each of the three filtration channels. A grid resolution of $g$ means that each channel is sampled at $g+1$ evenly spaced threshold values, so the resulting descriptor is a $(g+1)\times(g+1)\times(g+1)$ array, flattened into a single feature vector. Increasing $g$ therefore refines the sampling of the threshold space along all three filtration channels simultaneously, at the cost of a cubic growth in the number of features.

For PH, persistence diagrams are converted into persistence images by discretizing the birth-persistence plane, restricted to the fixed range $(0,1.25)\times(0,1.25)$, into an $N\times N$ pixel grid, over which the (weighted) persistence surface is evaluated. As a reminder, thecCone-filtration values assigned to solid voxels lie in $[0,1]$, whereas void voxels are assigned the value $1.25$, ensuring that they enter the filtration only after all solid voxels. Consequently, all finite birth and death values lie in $[0,1.25]$, and the corresponding persistence values are also bounded by $1.25$ which motivates the ranges chosen above. Increasing $N$ refines the resolution at which the birth-persistence plane is sampled, at the cost of a quadratic growth in the number of features per homological dimension.

\subsection{Results}

\begin{figure}[h]
    \centering
    \includegraphics[width=\textwidth]{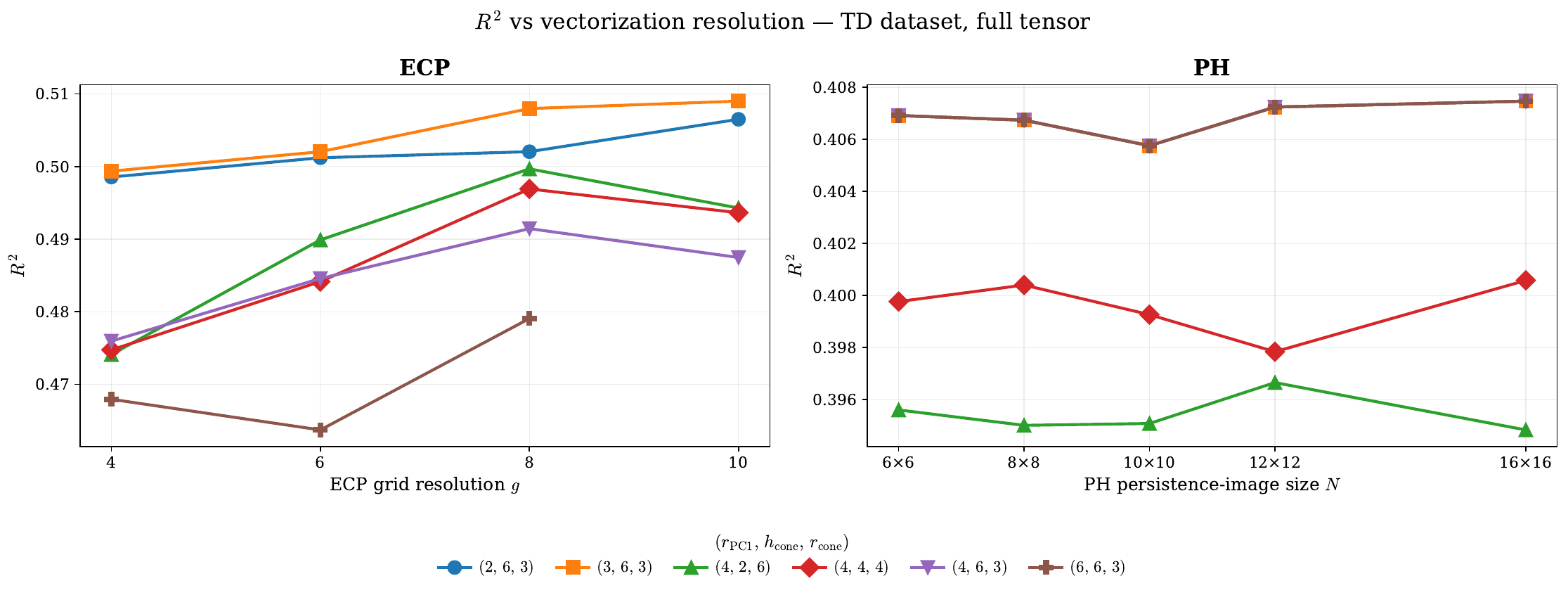}
    \caption{Cross-validated $R^2$ for full-tensor prediction on the TD dataset (all seven directions, $[100],[010],[001],[110],[101],[011],[111]$) as a function of the vectorization resolution, for Euler characteristic profiles (ECP, left) and persistent homology (PH, right). Each curve corresponds to one combination of filtration hyperparameters $r_{\mathrm{PC1}}$ (principal-component neighborhood radius), $h_{\mathrm{cone}}$ (cone height), and $r_{\mathrm{cone}}$ (cone base radius).}
    \label{fig:tda_hyperparam_scan}
\end{figure}

Figure~\ref{fig:tda_hyperparam_scan} shows $R^2$ as a function of the vectorization resolution ($g$ for ECP, $N$ for PH), with one curve per filtration hyperparameter combination described above. Across the full scan, $R^2$ varies only modestly: for ECP, all six filtration settings and all four grid resolutions fall within a narrow band, with $R^2$ increasing gradually with $g$ before leveling off around $g=8$; for PH, the dependence on both the vectorization resolution and the filtration hyperparameters is even weaker, with several of the curves nearly coinciding. This is a substantially smaller spread than the differences observed between fundamentally different descriptor families or between directional and non-directional descriptors elsewhere in this study, indicating that the exact choice of filtration and vectorization hyperparameters, within the ranges tested here, is not a major driver of predictive performance. The parameter combination used throughout the main text and applied unchanged to the RTP and ATTD datasets ($h_{\mathrm{cone}}=6$, $r_{\mathrm{cone}}=3$, $r_{\mathrm{PC1}}=3$, ECP grid resolution $g=8$, PH persistence-image size $N=12$) was chosen as a near-optimal, computationally cheaper trade-off: it lies within 0.001 of the best point found in the scan for both ECP ($g=10$) and PH ($N=16$), while requiring substantially fewer features.

\newpage
\section{Directional Baseline Descriptors}
\label{sec:baseline_descriptors}

To determine whether the predictive advantage of the direction-aware topological descriptors described in the main text arises specifically from their topological content or, more trivially, from the introduction of directional information in any form, three non-topological baseline descriptors were implemented: a \emph{porosity profile}, a \emph{two-point correlation function}, and a \emph{fabric tensor profile}. Those implementations are available in the project repository. In contrast to conventional usage, where each of these quantities is typically reported as a single global scalar or tensor for the whole structure, all three baselines were here made \emph{directional} in a manner directly analogous to the direction-aware TDA descriptors: each was computed independently for a fixed set of $L=7$ integer lattice  directions (similar to Miller indices) written $[hkl]$ for the (unnormalized) direction vector $(h,k,l)$,
$$
\hat{d} \in \big\{ [100],\,[010],\,[001],\,[110],\,[101],\,[011],\,[111] \big\},
$$                                                                                spanning the three axis-aligned directions ($[100]$, $[010]$, $[001]$, i.e., the grid's Cartesian axes), three face-diagonal directions ($[110]$, $[101]$, $[011]$), and one body-diagonal direction ($[111]$) of the cubic voxel grid.
This is the same set of directions used for the direction-aware TDA descriptors, allowing a like-for-like comparison at equal directional resolution. For each structure, the resulting $7$ per-direction profiles were concatenated into a single feature vector and passed to the same CatBoost gradient-boosted-tree pipeline used for the topological descriptors (see main text, Section~3), so that any performance difference between baselines and TDA descriptors reflects the descriptors themselves rather than the downstream model.

\subsection{Porosity profile}

For a direction $\hat{d}=(d_x,d_y,d_z)$, every voxel of the binary indicator field $X(\mathbf{x})$ is projected onto $\hat{d}$, $p(\mathbf{x}) = \hat{d}\cdot\mathbf{x}$, and voxels are grouped into slabs of constant (rounded) projection value, i.e., discrete layers perpendicular to $\hat{d}$. For axis-aligned directions these slabs coincide exactly with the grid planes; for face- and body-diagonal directions are tilted planes cutting across the grid.
The porosity profile is the mean solid fraction within each slab,
$$
P_{\hat d}(r) = \frac{1}{N_r}\sum_{\mathbf{x}\,:\,p(\mathbf{x})=r} X(\mathbf{x}),
$$
resampled by linear interpolation to $n_{\mathrm{out}}=80$ evenly spaced positions along $\hat d$, giving $7\times80=560$ features per structure. The porosity profile captures spatial variation of the local volume fraction along the loading direction — for example, layering, gradients, or periodic density modulation — and is directly related to the density-dependent scaling of stiffness that underlies classical structure-property relations such as the Gibson-Ashby model~\cite{Gibson1982} and to volume-fraction-based statistical descriptors of heterogeneous media~\cite{Torquato2002}.

\subsection{Two-point correlation function}

The two-point correlation function (also denoted $S_2$) is a standard statistical descriptor of heterogeneous and porous media~\cite{Torquato2002,PhysRevE.76.031110}, defined for a lag $r$ and direction $\hat d$ as the probability that two voxels separated by $r\hat d$ are both solid,
\[
S_2(r,\hat d) = P\big[X(\mathbf{x})=1 \ \text{and}\ X(\mathbf{x}+r\hat d)=1\big] = \big\langle X(\mathbf{x})\,X(\mathbf{x}+r\hat d) \big\rangle_{\mathbf{x}}.
\]
For each direction $(h,k,l)$, $S_2$ is evaluated at integer multiples of the step vector $(h,k,l)$ using periodic voxel shifts, which exploits the periodic boundary conditions of the structures and, for axis-aligned directions, reduces exactly to the conventional one-dimensional $S_2$ function. The sampled lag range is fixed at $[0,N/2]$ voxels (half of the domain size $N$, the maximum lag meaningful under periodicity) and resampled to $n_{\mathrm{out}}=41$ evenly spaced values, giving $7\times41=287$ features per structure. By construction $S_2(0,\hat d)$ equals the (direction-independent) solid volume fraction.

\subsection{Fabric tensor profile}

The fabric tensor is a classical second-order descriptor of the orientation distribution of solid-void interfaces (or particle/pore orientations), widely used to link microstructural anisotropy to elastic anisotropy in granular materials and trabecular bone~\cite{Kanatani1984,Cowin1985}. Rather than reducing the interface-orientation distribution to a single global tensor, a directional profile analogous to the porosity profile above was computed. At every solid-void interface voxel, the local unit surface normal $\hat n(\mathbf{x})$ is estimated from the normalized gradient of the binary indicator field, $\hat n = \nabla X / |\nabla X|$. For a direction $\hat d$, voxels are again grouped into slabs perpendicular to $\hat d$ (as for the porosity profile), and the fabric profile is defined as the slab-averaged squared alignment between interface normals and the direction of interest,
\[
F_{\hat d}(r) = \big\langle (\hat n(\mathbf{x}) \cdot \hat d)^2 \big\rangle_{\mathbf{x}\in\text{boundary},\, p(\mathbf{x})=r},
\]
resampled to $n_{\mathrm{out}}=80$ points, giving $7\times80=560$ features per structure. Values $F_{\hat d}(r)\approx 1$ indicate that interfaces within slab $r$ are predominantly oriented perpendicular to $\hat d$ (i.e., facing the loading direction), $F_{\hat d}(r)\approx 0$ indicates interfaces aligned parallel to $\hat d$, and $F_{\hat d}(r)\approx 1/3$ corresponds to a locally isotropic distribution of interface normals. Since the classical fabric-elasticity relations predict that the effective stiffness along a direction increases with the degree of interface alignment along it~\cite{Cowin1985}, the fabric profile provides a physically motivated, non-topological baseline that is directly comparable in spirit to the direction-aware topological descriptors, while relying only on local surface-orientation statistics rather than multi-scale connectivity information.

\newpage
\section{Additional metrics and results}
The raw per-component distributions underlying the summary statistics in this section, compared across RTP, TD, and ATTD, are shown in Section~\ref{sec:voigt_distributions}.
Tables~\ref{tab:mae_rtp} and \ref{tab:mae_td} report the mean absolute error (MAE, in GPa) for the same methods, target groups, and datasets (RTP and TD, respectively) as Tables~1 and~2 of the main text, complementing the $R^2$ values reported there.

\begin{table}[h]
\centering
\caption{Prediction Mean Absolute Error MAE across the 5 CV folds for each stiffness-tensor target group on the RTP dataset. Number in given in the parenthesis is the standard deviation of MAE across folds. For each column, the best result is shown in \textbf{bold} and the second-best in \textit{italic}.}
\label{tab:mae_rtp}
\begin{tabular}{l|rrrrr}
\multicolumn{1}{l}{Method} & \multicolumn{1}{c}{Uniaxial} & \multicolumn{1}{c}{Poisson} & \multicolumn{1}{c}{Shear} & \multicolumn{1}{c}{Off-diagonal} & \multicolumn{1}{c}{Full-Tensor} \\ \hline \hline
Porosity & 1.79(.16) & 0.69(.06) & 0.54(.06) & 0.62(.06) & 0.79(.07) \\
TPC & \textit{1.49(.18)} & \textbf{0.61(.07)} & 0.45(.05) & 0.59(.05) & \textit{0.72(.07)} \\
Fabric & 13.18(.53) & 6.01(.24) & 3.71(.14) & \textit{0.47(.04)} & 3.58(.14) \\
CNN & 1.92(.37) & 0.72(.12) & \textit{0.41(.05)} & 0.59(.03) & 0.87(.06) \\ \hline
TDA (undir.) & 2.59(.21) & 1.05(.09) & 0.73(.06) & 0.62(.06) & 0.99(.07) \\
TDA & \textbf{1.47(.17)} & \textit{0.64(.10)} & \textbf{0.40(.06)} & \textbf{0.36(.03)} & \textbf{0.58(.06)}
\end{tabular}
\end{table}

\begin{table}[h]
\centering
\caption{The same as Table \ref{tab:mae_rtp} but for the TD dataset.}
\label{tab:mae_td}
\begin{tabular}{l|rrrrr}
\multicolumn{1}{l}{Method} & \multicolumn{1}{c}{Uniaxial} & \multicolumn{1}{c}{Poisson} & \multicolumn{1}{c}{Shear} & \multicolumn{1}{c}{Off-diagonal} & \multicolumn{1}{c}{Full-Tensor} \\ \hline \hline
Porosity & 1.27(.06) & 0.48(.02) & 0.30(.01) & \textbf{0.11(.01)} & 0.36(.01) \\
TPC & 0.89(.02) & 0.41(.02) & 0.26(.01) & \textit{0.11(.01)} & \textit{0.29(.00)} \\
Fabric & 3.88(.17) & 1.28(.05) & 1.24(.05) & 0.12(.00) & 0.98(.04) \\
CNN & \textbf{0.70(.12)} & \textbf{0.28(.03)} & \textbf{0.24(.06)} & 0.14(.01) & 0.31(.07) \\ \hline
TDA (undir.) & 1.65(.07) & 0.49(.02) & 0.40(.02) & 0.13(.00) & 0.44(.01) \\
TDA & \textit{0.85(.02)} & \textit{0.33(.01)} & \textit{0.26(.01)} & 0.12(.00) & \textbf{0.27(.01)}
\end{tabular}
\end{table}

Because MAE is a scale-dependent metric, expressed directly in GPa rather than as a fraction of explained
variance, the absolute numbers differ from the $R^2$ picture, but the overall conclusions are essentially
unchanged. Directional TDA remains the best or second-best method on almost every target group in both
tables, consistently outperforming the non-directional baselines (Porosity, TPC, Fabric) and its own
undirectional counterpart, TDA (undir.). As with $R^2$, the improvement is most pronounced on the
Off-diagonal group -- and, by extension, on the pooled Full-Tensor group -- where TDA attains the lowest
MAE on RTP by a wide margin (0.36 GPa, versus 0.47 GPa for the next-best method, Fabric) and the lowest
Full-Tensor MAE on both datasets (0.58 GPa on RTP, 0.27 GPa on TD). On TD, the Off-diagonal MAE gap
between methods is much narrower in absolute terms, since the true off-diagonal stiffness components are
close to zero for these near-isotropic structures; here MAE mainly reflects this small intrinsic scale
rather than predictive skill, which is why the corresponding $R^2$ values (Table~2) remain the more
informative comparison for this target group. Crucially, TDA's advantage on the coupling terms is not
obtained at the expense of the diagonal target groups: on Uniaxial, Poisson, and Shear, TDA stays close
to the best-performing method on both datasets, ranking first or second throughout. The corresponding MAE
results for the ATTD dataset are reported in Table~\ref{tab:metrics_attd_mae} (Section~\ref{sec:results_attd}).

\newpage
\section{Results for the ATTD dataset}
\label{sec:results_attd}
The Anisotropic Transformed Topologically Diverse (ATTD) dataset is derived from the Topologically Diverse (TD) dataset, which comprises 2,376 structures spanning five families (Voronoi, zeolitic, diamond-like, cubic-strut, and spline-based) and is statistically isotropic, since anisotropy cannot be introduced in a controlled way at the level of its microstructure generation. To obtain a benchmark with tunable anisotropy while preserving this structural diversity, each TD structure was subjected to a simple geometric elongation procedure. The original $80\times80\times80$ binary voxel structure was first downsampled to a $40\times40\times40$ coarse-grained segment, which was then elongated along the $z$ direction to a $40\times40\times80$ volume and replicated four times in the transverse directions to reconstruct a periodic $80\times80\times80$ grid, yielding 4{,}750 structures in total. This transformation preserves the local topology of the parent structure while breaking isotropy and introducing a moderate, controllable elongation along $z$. As with TD, the solid phase is assigned the elastic properties of aluminium ($E_{\mathrm{bulk}}=70.0\,\mathrm{GPa}$, $\nu_{\mathrm{bulk}}=0.33$). ATTD thus serves as a complementary benchmark bridging the statistically isotropic TD dataset and the strongly anisotropic RTP dataset, letting us test whether the advantage of direction-aware topological descriptors over non-directional ones -- and their ability to close the gap with CNN baselines -- scales with the degree of anisotropy rather than being an artifact specific to the RTP generation process.
Section~\ref{sec:voigt_distributions} compares the resulting stiffness-tensor component distributions directly across all three datasets.

\begin{table}[h]
\centering
\caption{The same as Table 1 and Table 2 in the main text, but for the ATTD dataset.}
\label{tab:metrics_attd_r2}
\begin{tabular}{l|rrrrr}
\multicolumn{1}{l}{Method} & \multicolumn{1}{c}{Uniaxial} & \multicolumn{1}{c}{Poisson} & \multicolumn{1}{c}{Shear} & \multicolumn{1}{c}{Off-diagonal} & \multicolumn{1}{c}{Full-Tensor} \\ \hline \hline
Porosity & 0.914(.006) & 0.900(.007) & 0.947(.003) & -0.004(.003) & 0.388(.006) \\
TPC & 0.953(.003) & 0.926(.006) & \textit{0.960(.003)} & 0.041(.018) & \textit{0.405(.004)} \\
Fabric & 0.442(.020) & 0.361(.024) & 0.443(.027) & \textit{0.071(.004)} & 0.210(.008) \\
CNN & \textbf{0.981(.011)} & \textbf{0.967(.006)} & \textbf{0.977(.006)} & -0.015(.017) & 0.314(.026) \\ \hline
TDA (undir.) & 0.867(.004) & 0.874(.004) & 0.908(.005) & -0.005(.004) & 0.355(.015) \\
TDA & \textit{0.956(.008)} & \textit{0.928(.013)} & 0.955(.005) & \textbf{0.200(.016)} & \textbf{0.501(.013)}
\end{tabular}
\end{table}

\begin{table}[h]
\centering
\caption{The same as Table~\ref{tab:mae_rtp} but for the ATTD dataset.}
\label{tab:metrics_attd_mae}
\begin{tabular}{l|rrrrr}
\multicolumn{1}{l}{Method} & \multicolumn{1}{c}{Uniaxial} & \multicolumn{1}{c}{Poisson} & \multicolumn{1}{c}{Shear} & \multicolumn{1}{c}{Off-diagonal} & \multicolumn{1}{c}{Full-Tensor} \\ \hline \hline
Porosity & 1.36(.06) & 0.50(.01) & 0.36(.01) & 0.13(.00) & 0.40(.01) \\
TPC & 1.02(.05) & 0.41(.01) & 0.31(.01) & 0.13(.00) & 0.33(.01) \\
Fabric & 4.49(.07) & 1.43(.03) & 1.36(.01) & 0.13(.00) & 1.12(.02) \\
CNN & \textbf{0.76(.23)} & \textbf{0.30(.03)} & \textbf{0.27(.05)} & 0.13(.01) & \textbf{0.27(.01)} \\ \hline
TDA (undir.) & 1.79(.08) & 0.52(.01) & 0.45(.02) & 0.13(.00) & 0.47(.02) \\
TDA & \textit{0.96(.03)} & \textit{0.35(.02)} & \textit{0.30(.02)} & \textbf{0.12(.00)} & \textit{0.30(.01)}
\end{tabular}
\end{table}

\begin{figure}
    \centering
    \includegraphics[width=\linewidth]{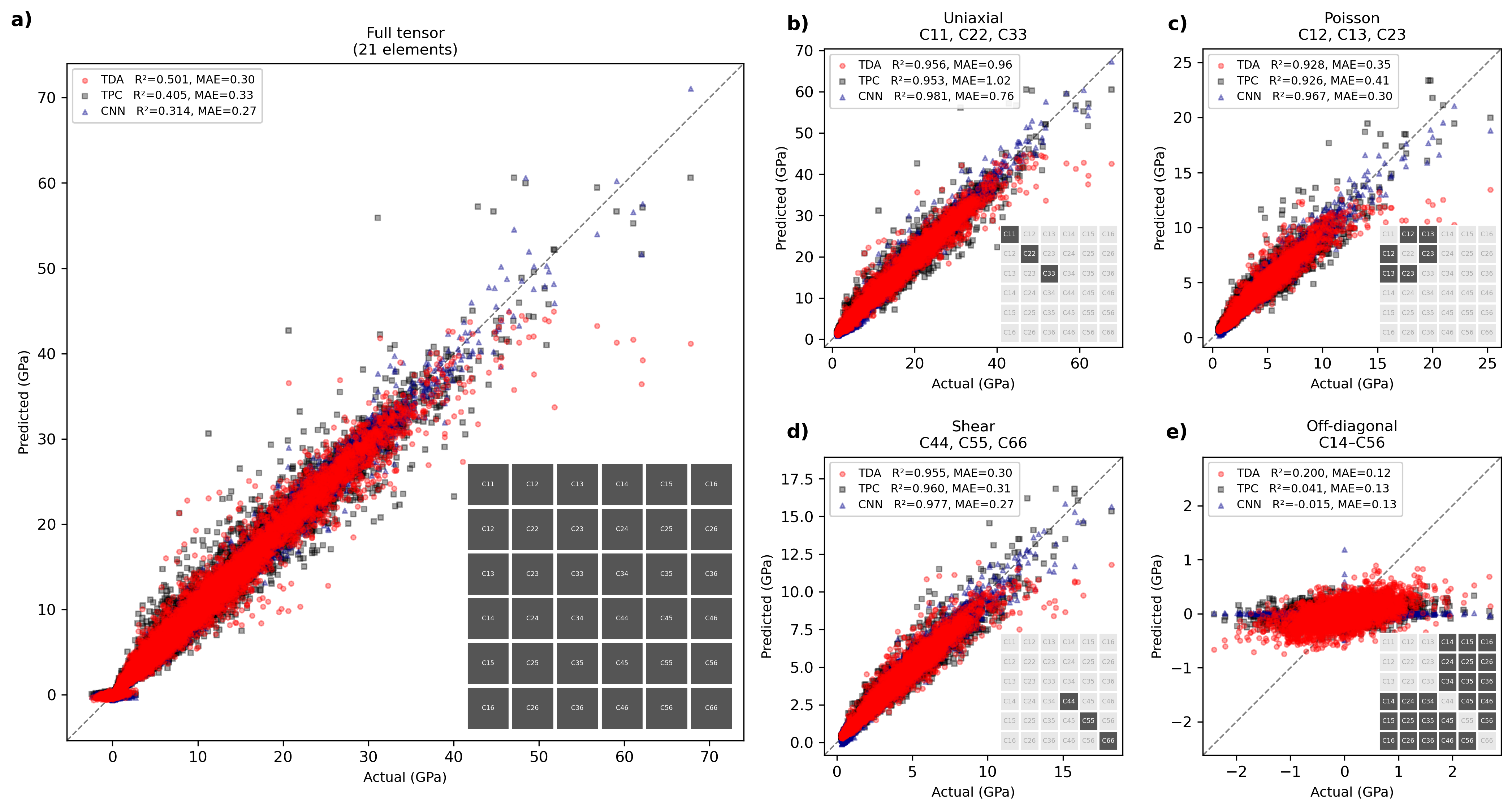}
    \caption{
    Predicted vs. actual stiffness-tensor components on the ATTD dataset, comparing TDA (red circles), TPC (black squares), and CNN (dark-blue triangles). Each panel pools test-set predictions from all 5 cross-validation folds; the dashed line marks perfect prediction. $R^2$ and MAE is averaged across the 5 CV folds. The inset in the lower-right corner of each panel is a schematic of the 6x6 Voigt stiffness matrix, with the independent elements belonging to that panel's target group shaded dark gray. a) Full tensor: all 21 independent elements pooled together. b) Uniaxial group: diagonal normal-stress terms $C_{11}$, $C_{22}$, $C_{33}$. c) Poisson group: off-diagonal normal-normal coupling terms $C_{12}$, $C_{13}$, $C_{23}$. d) Shear group: diagonal shear terms $C_{44}$, $C_{55}$, $C_{66}$. e) Off-diagonal group: the 12 normal-shear and shear-shear coupling terms.
    }
    \label{fig:scatter_attd}
\end{figure}

\newpage
\section{Statistical analysis}

The $R^2$/MAE tables report a single mean over the whole test set, which can mask outlier-driven effects and cannot show whether one
method is consistently more accurate than another on the same structures. We therefore also test, structure by structure, whether
directional TDA's error is systematically lower than each comparison method's.

For each (dataset, target group, method) combination, every test structure $i$ is assigned a per-structure error, defined as the mean
absolute error over the Voigt components belonging to that target group. Structures are aligned across methods (all methods are
evaluated on the same structures under the same cross-validation folds) and pooled across all 5 folds.

For each matched pair we compute $d_i = \mathrm{err}_{\mathrm{TDA}}(i) - \mathrm{err}_{\mathrm{comparison}}(i)$ and apply a one-sided
Wilcoxon signed-rank test~\cite{montgomery2018, devore2020}:
\begin{itemize}
    \item[] $H_0$: $d_i$ is symmetric about zero (neither method is systematically more accurate).
    \item[] $H_1$: the distribution of $d_i$ is shifted below zero (TDA is systematically more accurate).
\end{itemize}
The signed-rank test is used instead of a paired $t$-test because per-structure errors are strictly positive and right-skewed, violating
the normality assumption; the one-sided alternative reflects the directional hypothesis already suggested by the $R^2$/MAE tables. A two-sided test was also conducted for comparison: since the two-sided $p$-value for the Wilcoxon signed-rank test is $\min(1, 2p_{\text{one-sided}})$, and the one-sided $p$-values obtained here are overwhelmingly extreme (below $\sim 10^{-8}$), doubling them leaves every significance conclusion unchanged; the two-sided test therefore reproduces the same qualitative picture throughout.
Since
the median of $d_i$ can occasionally be positive even when TDA's mean error is lower (driven by a few large errors elsewhere), results
are additionally flagged whenever the comparison method has the lower median error, so a low p-value is never misread.

Each (dataset, target group, method) cell is tested independently, with no multiple-comparison correction: nearly all p-values are
either far below $p<0.01$ or far above $p=0.05$ (often $p=1$), so the qualitative pattern is insensitive to this choice; single- vs.\
double-asterisk marking lets a reader apply a stricter threshold if desired. 
 
A conservative Bonferroni correction across all 75 tests ($\alpha = 0.05/75 \approx 6.7\times10^{-4}$) leaves every Off-diagonal and Full-Tensor comparison significant (all $p \lesssim 10^{-20}$) -- the comparisons underlying the paper's central claim -- with only a small number of comparisons within the already closely-matched Uniaxial/Poisson/Shear groups (against TPC or CNN) falling below this stricter threshold, consistent with those groups being a near-tie between methods.

This aggregated test operates at the level of the five target groups (Uniaxial, Poisson, Shear, Off-diagonal, Full-Tensor), each using
the model trained specifically on that group, and compares directional TDA against Porosity, TPC, Fabric, CNN, and the
direction-agnostic TDA (undir.) descriptor. A complementary per-component analysis, isolating which individual Voigt components drive
each aggregated result, is discussed separately below.

\begin{table}[]
\centering
\caption{One-sided Wilcoxon signed-rank $p$-values for the RTP dataset, testing whether directional TDA's per-structure error is systematically
lower than that of each comparison method (rows), separately for each target group (columns). * denotes $p<0.05$ and ** denotes
$p<0.01$; cells without an asterisk are not significant at the 0.05 level. An upward arrow ($\uparrow$), where present, marks cases in
which the comparison method -- not TDA -- has the lower median error, i.e.\ the effect runs opposite to the tested direction.}
\label{tab:wilcoxon_rtp}
\begin{tabular}{l|lllll}
TDA vs & Uniaxial & Poisson & Shear & OffDiagonal & FullTensor \\ \hline
Porosity & 4.49e-16** & 7.20e-09** & 9.26e-25** & 1.54e-71** & 1.48e-50** \\
TPC & 6.69e-03** & 1.14e-03** & 1.07e-08** & 1.71e-71** & 6.71e-36** \\
Fabric & 9.69e-84** & 1.33e-82** & 9.59e-83** & 6.39e-45** & 7.35e-84** \\
CNN & 3.34e-20** & 1.39e-10** & 1.09e-02* & 7.53e-73** & 9.22e-62** \\
TDA (undir.) & 2.39e-51** & 6.34e-40** & 4.38e-58** & 1.29e-71** & 1.66e-73**
\end{tabular}
\end{table}
\begin{table}[]
\centering
\caption{The same as Table \ref{tab:wilcoxon_rtp} but for the TD dataset.}
\label{tab:wilcoxon_td}
\begin{tabular}{l|lllll}
TDA vs & Uniaxial & Poisson & Shear & OffDiagonal & FullTensor \\ \hline
Porosity & 1.05e-75** & 2.59e-96** & 6.44e-24** & 5.25e-46** & 2.37e-118** \\
TPC & 2.76e-02* & 2.07e-50** & 3.53e-02* & 1.82e-68** & 8.39e-26** \\
Fabric & 0.00e+00** & 9.49e-294** & 0.00e+00** & 7.52e-58** & 0.00e+00** \\
CNN & 1.00e+00 ↑ & 1.00e+00 ↑ & 7.34e-04** & 1.42e-113** & 4.45e-59** \\
TDA (undir.) & 3.47e-206** & 1.47e-102** & 6.15e-116** & 2.26e-45** & 9.86e-236**
\end{tabular}
\end{table}
\begin{table}[]
\centering
\caption{The same as Table \ref{tab:wilcoxon_rtp} but for the ATTD dataset.}
\label{tab:wilcoxon_attd}
\begin{tabular}{l|lllll}
TDA vs & Uniaxial & Poisson & Shear & OffDiagonal & FullTensor \\ \hline
Porosity & 5.21e-90** & 9.41e-89** & 1.90e-39** & 7.79e-28** & 3.45e-116** \\
TPC & 3.21e-04** & 2.32e-30** & 1.24e-03** & 7.74e-43** & 2.00e-22** \\
Fabric & 0.00e+00** & 0.00e+00** & 0.00e+00** & 2.32e-77** & 0.00e+00** \\
CNN & 1.00e+00 ↑ & 1.00e+00 ↑ & 4.65e-01 & 2.73e-67** & 1.00e+00 \\
TDA (undir.) & 6.66e-232** & 1.43e-111** & 1.81e-113** & 7.96e-28** & 7.32e-239**
\end{tabular}
\end{table}

Tables~\ref{tab:wilcoxon_rtp}--\ref{tab:wilcoxon_attd} confirm, at the level of individual structures, the ranking already suggested by
the aggregate $R^2$ (Table 1 and in the main text and Table \ref{tab:metrics_attd_r2}) MAE tables (Tables \ref{tab:mae_rtp}, \ref{tab:mae_td}, \ref{tab:metrics_attd_mae}). On RTP, directional TDA is significantly more accurate than all five other methods on every target
group ($p<0.05$ throughout, and $p<0.01$ for all but one cell), including the CNN. On TD and ATTD, TDA remains significantly better
than Porosity, Fabric, TPC, and the direction-agnostic TDA (undir.) descriptor across all five target groups. Its comparison with the
CNN is target-group-dependent: the CNN is more accurate than TDA on Uniaxial and Poisson (flagged by the reversed-effect
arrow), TDA is significantly more accurate on Off-diagonal (and, for TD, on Full-Tensor), and the Shear comparison, together with the
ATTD Full-Tensor comparison, is not statistically significant in either direction. In the latter case the median per-structure error
still nominally favors TDA, but the difference is far too small relative to the sample variability to be distinguished from chance --
unlike the Uniaxial/Poisson cells, where the effect is both large and significant in the opposite direction.

Taken together, these tests show that directional TDA's advantage over the non-CNN baselines is not an artifact of averaging over test
structures or cross-validation folds, but holds consistently structure by structure. Against the CNN, the significance pattern sharpens
the picture from the main text: the CNN's edge is confined to the diagonal target groups on the more topologically diverse TD/ATTD
datasets, while TDA's advantage on the off-diagonal (and, by inheritance, full-tensor) components is both large and robust across all
three datasets.

\newpage
\section{Stiffness Tensor Component Distributions}
\label{sec:voigt_distributions}

\begin{figure}[h]
    \centering
    \includegraphics[width=\textwidth]{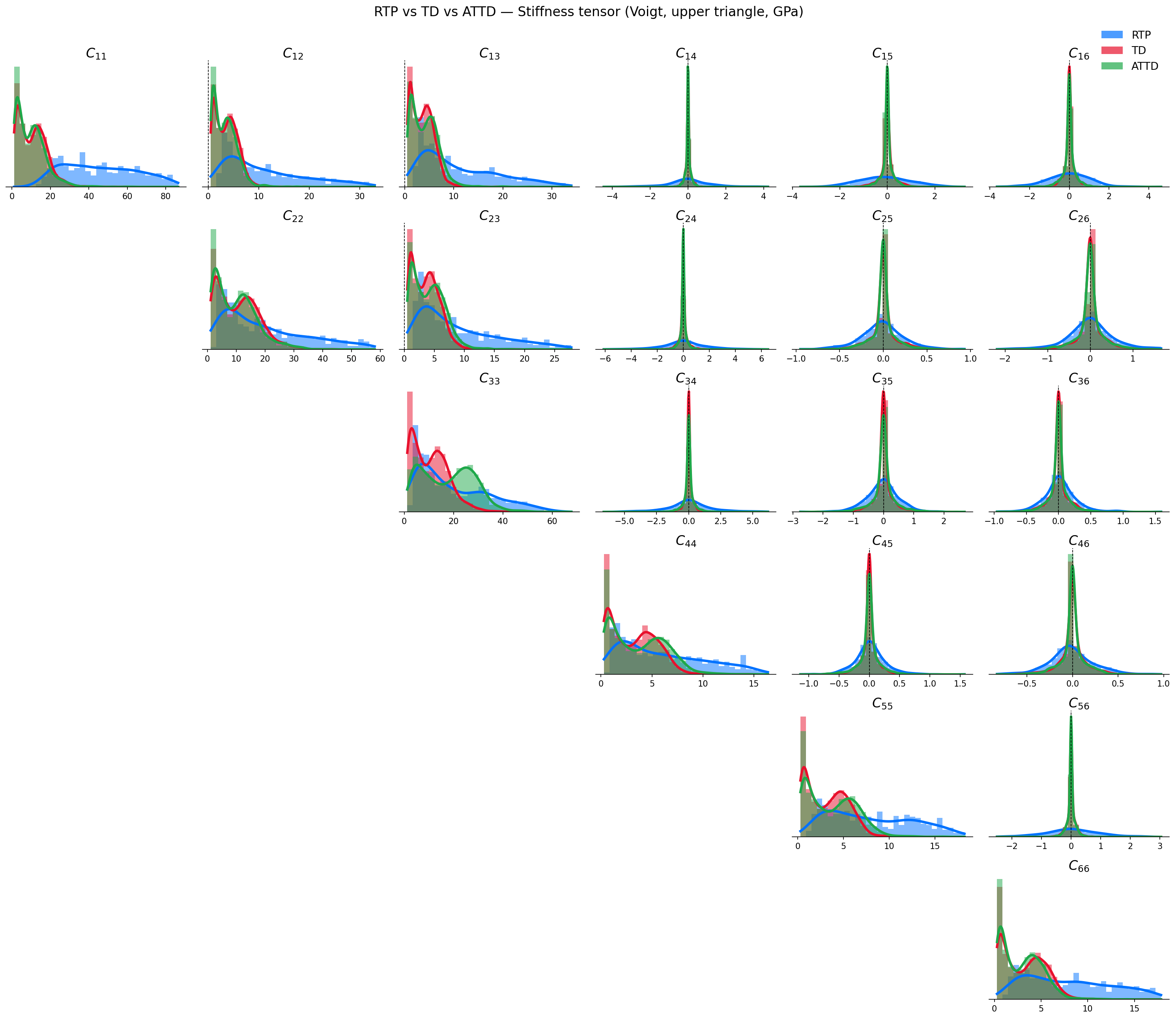}
    \caption{Distribution of the 21 independent stiffness-tensor components (Voigt notation, GPa) for the
    RTP, TD, and ATTD datasets. Each panel overlays alpha-blended histograms with a kernel density estimate
    (KDE) for each dataset; the dashed vertical line marks zero for the off-diagonal (coupling) components.}
    \label{fig:voigt_all_datasets}
\end{figure}

Figure~\ref{fig:voigt_all_datasets} shows the distribution of all 21 independent components of the
effective stiffness tensor $C_{ij}$ (Voigt notation, upper triangle) for the RTP, TD, and ATTD datasets
overlaid on the same axes. Differences between the three distributions reflect two compounded factors:
differences in structural topology (each dataset is generated by a different process, spanning different
ranges of porosity and pore-network anisotropy) and differences in the solid-phase material assigned to
each dataset (Au$_{0.30}$Ag$_{0.70}$, $E=81.5$\,GPa, for RTP; aluminium, $E=70.0$\,GPa, for TD and ATTD).
Consequently, the spread and location of these distributions should not be read as reflecting topology
alone.

The RTP distributions are visibly broader and shifted toward higher stiffness values than TD and ATTD
across nearly all diagonal components, consistent with both its stiffer constituent material and its
wider, deliberately constructed range of porosities and anisotropies; TD and ATTD, which share the same
base material and a common parent topology (ATTD being a moderately elongated version of TD), instead
show closely overlapping distributions throughout.

\section{Computational cost}
\label{sec:cost}

Since the direction-aware descriptors are proposed as a surrogate for, rather than an
improvement on, direct numerical homogenization, it is useful to state what evaluating a single
structure costs along each route. Table~\ref{tab:cost} reports measured timings for one
structure.

All timings are wall-clock times measured on a single core of an otherwise idle machine
(AMD EPYC 7402P, 12 physical cores and 24 threads), so that the two routes are compared under
identical conditions. Each entry is a mean over six structures of the RTP dataset. The
homogenization timing is for a single solver call, which returns the complete $6\times6$
stiffness tensor from six independent unit-strain load cases. The descriptor timings are quoted
per loading direction, and the complete descriptor uses seven directions. Inference with the
trained gradient-boosted model is negligible by comparison.

\begin{table}[h]
\centering
\caption{Measured cost of evaluating one structure on a single core. Descriptor steps are per
loading direction; the complete descriptor uses seven directions.}
\label{tab:cost}
\begin{tabular}{lr}
\hline
Step & Time (s) \\
\hline
FFT homogenization, six unit-strain load cases & 965 \\
\hline
Directional filtration (per direction)          & 1.2 \\
Euler characteristic profile (per direction)    & 1.9 \\
Persistent homology (per direction)             & 3.8 \\
Complete descriptor, seven directions           & 48 \\
\hline
\end{tabular}
\end{table}

Predicting the stiffness tensor from direction-aware topological descriptors is therefore
roughly a factor of $20$ cheaper per structure than computing it directly, and the same
descriptor is reused for all $21$ independent tensor components rather than being recomputed
per component. Repeating the measurement with all $24$ threads of the machine available to both
routes gives the same ratio, since both benefit from parallel execution to a similar degree.

The comparison above concerns the evaluation of a single structure. As with any surrogate model,
the time needed to assemble the training set and to fit the model has to be taken into account as
well. These are one-off costs amortized over all subsequent predictions, so
the approach is advantageous when the number of structures to be evaluated is large compared with
the size of the training set.

The convolutional network used for comparison was trained on an NVIDIA GeForce RTX~3090 with
$24$~GB of memory, taking approximately five minutes per cross-validation fold. This network is
applied directly to the voxelized structure and does not use the descriptors introduced here, so
its cost is not part of the comparison above. It is included as an additional reference point.

\clearpage
\section{Sensitivity of the reported metrics to the training configuration}
\label{sec:hyperparam_sensitivity}

The models trained on the descriptors compared in Tables~1 and~2 in the main
text are gradient-boosted decision-tree ensembles. Unlike a neural network, such a model contains no
weights that must be initialized: the trees are grown one after another by a deterministic greedy
procedure, and repeating an identical training run reproduces every reported metric exactly. The
metrics can nevertheless depend on how the data are partitioned and on the hyperparameters that
govern the training procedure. This section quantifies both.

All results in the manuscript were obtained with the library defaults: ensembles of $1000$ trees
of depth $6$, with the learning rate selected automatically by the library ($\approx 0.03$ for
the dataset sizes used here) and training stopped early on the validation fold. 

Table~\ref{tab:hyperparam_sensitivity} reports the effect of departing from this configuration.
The complete five-fold cross-validation was repeated, on the RTP dataset with the direction-aware
(TDA) descriptors, under seven additional configurations grouped in three families:

\begin{itemize}
\item \textbf{Data partition.} The random seed controlling the partition was changed. This
      reassigns structures to the five cross-validation folds and redraws the subsequent
      train/validation split, so each seed evaluates the model on a different partition of the
      dataset.
\item \textbf{Training protocol.} The learning rate, that is, the shrinkage factor applied to
      each successive tree before it is added to the ensemble, was set explicitly to $0.03$ and
      to $0.10$; and the number of boosting iterations, that is, the number of trees grown, was
      doubled from $1000$ to $2000$.
\item \textbf{Model capacity.} The depth of the individual trees, which is the principal capacity
      parameter of a boosted ensemble, was set to $4$ and to $8$.
\end{itemize}

\begin{table}[h]
\centering
\caption{Coefficient of determination $R^2$ on the RTP dataset for the direction-aware (TDA)
descriptors, obtained under variations of the data partition and of the training hyperparameters.
Values are means over the five cross-validation folds, with the standard deviation across folds.
The first row is the configuration reported in Table~1 in main text. 
Setting the learning
rate explicitly to $0.03$ reproduces that row, confirming the value selected automatically by the
library for this dataset.}
\label{tab:hyperparam_sensitivity}
\begin{tabular}{lccccc}
\hline
Configuration & Uniaxial & Poisson & Shear & Off-diagonal & Full-Tensor \\
\hline
Reported & $0.980 \pm 0.007$ & $0.980 \pm 0.007$ & $0.982 \pm 0.006$ & $0.468 \pm 0.019$ & $0.672 \pm 0.016$ \\
\hline
Partition seed 2 & $0.981 \pm 0.005$ & $0.982 \pm 0.006$ & $0.983 \pm 0.004$ & $0.476 \pm 0.029$ & $0.677 \pm 0.017$ \\
Partition seed 3 & $0.979 \pm 0.007$ & $0.980 \pm 0.008$ & $0.982 \pm 0.004$ & $0.468 \pm 0.016$ & $0.671 \pm 0.015$ \\
\hline
Learning rate $0.03$ & $0.980 \pm 0.007$ & $0.980 \pm 0.007$ & $0.982 \pm 0.006$ & $0.468 \pm 0.019$ & $0.672 \pm 0.016$ \\
Learning rate $0.10$ & $0.976 \pm 0.008$ & $0.979 \pm 0.005$ & $0.978 \pm 0.006$ & $0.447 \pm 0.027$ & $0.673 \pm 0.015$ \\
$2000$ boosting iterations & $0.980 \pm 0.007$ & $0.980 \pm 0.007$ & $0.982 \pm 0.006$ & $0.472 \pm 0.020$ & $0.678 \pm 0.016$ \\
\hline
Tree depth $4$ & $0.985 \pm 0.004$ & $0.986 \pm 0.004$ & $0.987 \pm 0.004$ & $0.502 \pm 0.012$ & $0.694 \pm 0.009$ \\
Tree depth $8$ & $0.974 \pm 0.008$ & $0.974 \pm 0.007$ & $0.976 \pm 0.007$ & $0.412 \pm 0.021$ & $0.636 \pm 0.019$ \\
\hline
\end{tabular}
\end{table}

Changing the data partition moves the three diagonal groups by at most $0.002$ and the
Off-diagonal group by $0.008$. The training protocol has a similarly limited effect: doubling the
number of iterations changes no group by more than $0.006$, since early stopping terminates
training before the additional trees are used, and a threefold increase of the learning rate costs
$0.021$ on the Off-diagonal group while leaving the remaining groups within $0.004$. Tree depth
is the most influential of the parameters examined, as expected of the capacity parameter: depth
$8$ lowers the Off-diagonal $R^2$ to $0.412$, and depth $4$ raises it to $0.502$, slightly above
the reported value. The latter difference is of the same order as the standard deviation across
folds and is obtained without any search over the remaining parameters; we therefore report the
default configuration throughout rather than the best one found here.

Across all configurations the three diagonal groups vary by at most $0.012$ in $R^2$, the
Off-diagonal group spans $0.412$--$0.502$ and the Full-Tensor group $0.636$--$0.694$. These
ranges are comparable to the fold-to-fold standard deviation of a single configuration, and are
substantially smaller than the differences between descriptor families reported in
Table~1 in the Manuscript, where the best-performing alternative descriptor attains
$R^2 = 0.287$ on the Off-diagonal group and $0.415$ on the Full-Tensor group. The ordering of the
methods, and the conclusions drawn from it, are therefore not sensitive to the choices examined here.

The convolutional network used as a baseline is a deep architecture and, unlike the boosted
models, does have a weight initialization stage: it is trained from scratch, the convolutional
weights being drawn from the He (Kaiming) normal distribution~\cite{He2015} with fan-out scaling, the
batch-normalization weights and biases set to one and zero, and the biases of the final linear
layer set to zero. We repeated its training three times in the reported configuration,
under two further random seeds, which change the initial weights and the data partition
together, and at
learning rates of $3 \times 10^{-4}$ and $3 \times 10^{-3}$ in place of the $10^{-3}$ used in main text. Across these seven runs the diagonal groups remain within $0.963$--$0.987$, while the
Off-diagonal group varies between $-0.154$ and $+0.003$ and the Full-Tensor group between $0.131$
and $0.343$. This variation is of the same order as the standard deviation across folds already
reported for the network in Table~1 in main text, and the lower of the two learning rates
improves it slightly over the reported configuration on the Full-Tensor group ($0.343$ against
$0.280$). None of the configurations alters the comparison: the network remains indistinguishable
from zero on the Off-diagonal group throughout, and no configuration approaches the values
obtained with the direction-aware descriptors on the two groups that separate the methods.

\bibliographystyle{sn-mathphys-num}
\bibliography{references}

@article{
ShanShi,
author = {Shan Shi  and Yong Li  and Bao-Nam Ngo-Dinh  and Jürgen Markmann  and Jörg Weissmüller },
title = {Scaling behavior of stiffness and strength of hierarchical network nanomaterials},
journal = {Science},
volume = {371},
number = {6533},
pages = {1026-1033},
year = {2021},
doi = {10.1126/science.abd9391},
URL = {https://www.science.org/doi/abs/10.1126/science.abd9391},
eprint = {https://www.science.org/doi/pdf/10.1126/science.abd9391}}

@article{praca_z_madrytem,
title = {Transferable 3D convolutional neural networks for elastic constants prediction in nanoporous metals},
journal = {Materials \& Design},
volume = {260},
pages = {114896},
year = {2025},
issn = {0264-1275},
doi = {https://doi.org/10.1016/j.matdes.2025.114896},
url = {https://www.sciencedirect.com/science/article/pii/S0264127525013164},
author = {Sergei Zorkaltsev and Rafał Topolnicki and Tal-El Carmon and Santhosh Mathesan and Paweł Dłotko and Dan Mordehai and Maciej Haranczyk}
}

@article{ZANDERSONS2021116979,
title = {On factors defining the mechanical behavior of nanoporous gold},
journal = {Acta Materialia},
volume = {215},
pages = {116979},
year = {2021},
issn = {1359-6454},
doi = {https://doi.org/10.1016/j.actamat.2021.116979},
url = {https://www.sciencedirect.com/science/article/pii/S1359645421003591},
author = {Birthe Zandersons and Lukas Lührs and Yong Li and Jörg Weissmüller}
}

@misc{Prokhorenkova2019,
      title={CatBoost: unbiased boosting with categorical features}, 
      author={Liudmila Prokhorenkova and Gleb Gusev and Aleksandr Vorobev and Anna Veronika Dorogush and Andrey Gulin},
      year={2019},
      eprint={1706.09516},
      archivePrefix={arXiv},
      primaryClass={cs.LG},
      url={https://arxiv.org/abs/1706.09516}, 
}

@article{BEETS2021116445,
title = {The Mechanical Response of Nanoporous Gold and Silver Foams with Varying Composition and Surface Segregation},
journal = {Acta Materialia},
volume = {203},
pages = {116445},
year = {2021},
issn = {1359-6454},
doi = {https://doi.org/10.1016/j.actamat.2020.10.064},
url = {https://www.sciencedirect.com/science/article/pii/S1359645420308624},
author = {Nathan Beets and Diana Farkas and Karsten Albe}
}

@misc{wolframBSplineFunction,
   author = {Wolfram Research},
   journal = {Wolfram Research (2008), BSplineFunction, Wolfram Language function, https://reference.wolfram.com/language/ref/BSplineFunction.html},
   title = {BSplineFunction, Wolfram Language function—Wolfram Documentation},
   url = {https://reference.wolfram.com/language/ref/BSplineFunction.html},
   year = {2008}
}

@book{deBoor1978,
  author    = {de Boor, Carl},
  title     = {A Practical Guide to Splines},
  edition   = {Revised},
  series    = {Applied Mathematical Sciences},
  volume    = {27},
  publisher = {Springer},
  address   = {New York, NY},
  year      = {2001},
  isbn      = {978-0-387-95366-3},
  url       = {https://link.springer.com/book/9780387953663}
}

@article{Lucarini2019,
   author = {S. Lucarini and J. Segurado},
   doi = {10.1007/S00466-018-1598-1/FIGURES/9},
   issn = {01787675},
   issue = {2},
   journal = {Computational Mechanics},
   month = {2},
   pages = {365-382},
   publisher = {Springer Verlag},
   title = {On the accuracy of spectral solvers for micromechanics based fatigue modeling},
   volume = {63},
   url = {https://link.springer.com/article/10.1007/s00466-018-1598-1},
   year = {2019}
}

@article{Zheng2014,
   author = {Xiaoyu Zheng and Howon Lee and Todd H. Weisgraber and Maxim Shusteff and Joshua DeOtte and Eric B. Duoss and Joshua D. Kuntz and Monika M. Biener and Qi Ge and Julie A. Jackson and Sergei O. Kucheyev and Nicholas X. Fang and Christopher M. Spadaccini},
   doi = {10.1126/SCIENCE.1252291},
   issn = {10959203},
   issue = {6190},
   journal = {Science},
   month = {6},
   pages = {1373-1377},
   pmid = {24948733},
   publisher = {American Association for the Advancement of Science},
   title = {Ultralight, ultrastiff mechanical metamaterials},
   volume = {344},
   url = {https://www.science.org/doi/10.1126/science.1252291},
   year = {2014}
}

@article{PhysRevE.76.031110,
  title = {Modeling heterogeneous materials via two-point correlation functions: Basic principles},
  author = {Jiao, Y. and Stillinger, F. H. and Torquato, S.},
  journal = {Phys. Rev. E},
  volume = {76},
  issue = {3},
  pages = {031110},
  numpages = {15},
  year = {2007},
  month = {Sep},
  publisher = {American Physical Society},
  doi = {10.1103/PhysRevE.76.031110},
  url = {https://link.aps.org/doi/10.1103/PhysRevE.76.031110}
}

@article{Badwe2017,
   author = {Nilesh Badwe and Xiying Chen and Karl Sieradzki},
   doi = {10.1016/J.ACTAMAT.2017.02.040},
   issn = {1359-6454},
   journal = {Acta Materialia},
   month = {5},
   pages = {251-258},
   publisher = {Pergamon},
   title = {Mechanical properties of nanoporous gold in tension},
   volume = {129},
   year = {2017}
}

@article{Gibson1982,
   author = {L. J. Gibson and M. F. Ashby},
   doi = {10.1098/RSPA.1982.0088},
   issn = {0080-4630},
   issue = {1782},
   journal = {Proceedings of the Royal Society of London. A. Mathematical and Physical Sciences},
   month = {7},
   pages = {43-59},
   publisher = {The Royal Society},
   title = {The mechanics of three-dimensional cellular materials},
   volume = {382},
   url = {https://dx.doi.org/10.1098/rspa.1982.0088},
   year = {1982}
}

@article{Robins2011,
   author = {Vanessa Robins and Peter John Wood and Adrian P. Sheppard},
   doi = {10.1109/TPAMI.2011.95},
   issn = {01628828},
   issue = {8},
   journal = {IEEE Transactions on Pattern Analysis and Machine Intelligence},
   pages = {1646-1658},
   title = {Theory and algorithms for constructing discrete morse complexes from grayscale digital images},
   volume = {33},
   year = {2011}
}

@article{Dlotko2024,
   author = {Paweł Dłotko},
   doi = {10.4171/MAG/190},
   issn = {2747-7894},
   issue = {132},
   journal = {European Mathematical Society Magazine},
   month = {6},
   pages = {5-13},
   publisher = {EMS Press},
   title = {On the shape that matters – topology and geometry in data science},
   url = {https://euromathsoc.org/magazine/articles/190},
   year = {2024}
}

@misc{Gurnari2025,
   author = {Davide Gurnari},
   publisher = {University of Warsaw},
   title = {New Shape Descriptors for Topological Data Analysis},
   url = {https://repozytorium.uw.edu.pl//handle/item/166201},
   year = {2025}
}

@article{Depboylu2024,
   author = {Fatma Nur Depboylu and Beliz Taşkonak and Petek Korkusuz and Evren Yasa and Olatunji Ajiteru and Kyu and Young Choi and Chan and Hum Park and Özgür Poyraz and Andrei-Alexandru Popa and Feza Korkusuz},
   doi = {10.1007/s42247-024-00774-2},
   isbn = {0123456789},
   journal={Emergent Materials},
   pages = {2711-2729},
   title = {Cleaning and coating procedures determine biological properties of gyroid porous titanium implants},
   volume = {7},
   url = {https://doi.org/10.1007/s42247-024-00774-2},
   year = {2024}
}

@article{Matassi2013,
   author = {Fabrizio Matassi and Alessandra Botti and Luigi Sirleo and Christian Carulli and Massimo Innocenti},
   doi = {10.11138/ccmbm/2013.10.2.111},
   issn = {17248914},
   issue = {2},
   journal = {Clinical Cases in Mineral and Bone Metabolism},
   month = {5},
   pages = {111},
   pmid = {24133527},
   title = {Porous metal for orthopedics implants},
   volume = {10},
   url = {https://pmc.ncbi.nlm.nih.gov/articles/PMC3796997/},
   year = {2013}
}

@article{Guo2025,
   author = {Chunliang Guo and Tao Ding and Yuan Cheng and Jianqing Zheng and Xiule Fang and Zhiyun Feng},
   doi = {10.3389/FBIOE.2025.1548675},
   issn = {22964185},
   journal = {Frontiers in Bioengineering and Biotechnology},
   pages = {1548675},
   pmid = {40078794},
   publisher = {Frontiers Media SA},
   title = {The rational design, biofunctionalization and biological properties of orthopedic porous titanium implants: a review},
   volume = {13},
   url = {https://pmc.ncbi.nlm.nih.gov/articles/PMC11897010/},
   year = {2025}
}

@article{Koya2021,
   author = {Alemayehu Nana Koya and Xiangchao Zhu and Nareg Ohannesian and A Ali Yanik and Alessandro Alabastri and Remo Proietti Zaccaria and Roman Krahne and Wei-Chuan Shih and Denis Garoli},
   doi = {10.1021/acsnano.0c10945},
   journal = {ACS Nano},
   title = {Nanoporous Metals: From Plasmonic Properties to Applications in Enhanced Spectroscopy and Photocatalysis},
   volume = {15},
   url = {www.acsnano.org},
   year = {2021}
}

@misc{deem2023pcod,
   author = {Michael W Deem},
   doi = {10.5281/ZENODO.8035241},
   title = {Michael Deem's PCOD database of Predicted Zeolitic Structures},
   url = {https://zenodo.org/records/8035241}
}

@article{C0CP02255A,
   author = {Ramdas Pophale and Phillip A. Cheeseman and Michael W. Deem},
   doi = {10.1039/C0CP02255A},
   issn = {14639076},
   issue = {27},
   journal = {Physical Chemistry Chemical Physics},
   pages = {12407-12412},
   pmid = {21423937},
   title = {A database of new zeolite-like materials},
   volume = {13},
   year = {2011}
}

@article{bresenham1965algorithm,
   author = {J. E. Bresenham},
   issue = {4},
   journal = {IBM Systems Journal},
   pages = {25-30},
   title = {Algorithm for computer control of a digital plotter},
   volume = {1},
   url = {https://sci-hub.se/10.1147/sj.41.0025},
   year = {1965}
}

@article{Lisitsa2020,
   author = {Vadim Lisitsa and Yaroslav Bazaikin and Tatyana Khachkova},
   doi = {10.1016/J.APM.2020.06.037},
   issn = {0307-904X},
   journal = {Applied Mathematical Modelling},
   month = {12},
   pages = {21-37},
   publisher = {Elsevier},
   title = {Computational topology-based characterization of pore space changes due to chemical dissolution of rocks},
   volume = {88},
   year = {2020}
}

@article{Thompson2023,
   author = {E. P. Thompson and B. R. Ellis},
   doi = {10.1029/2023WR034559},
   issn = {1944-7973},
   issue = {9},
   journal = {Water Resources Research},
   month = {9},
   pages = {e2023WR034559},
   publisher = {John Wiley \& Sons, Ltd},
   title = {Persistent Homology as a Heterogeneity Metric for Predicting Pore Size Change in Dissolving Carbonates},
   volume = {59},
   url = {https://agupubs.onlinelibrary.wiley.com/doi/10.1029/2023WR034559},
   year = {2023}
}

@article{Moon2019,
   author = {Chul Moon and Scott A. Mitchell and Jason E. Heath and Matthew Andrew},
   doi = {10.1029/2019WR025171},
   issn = {1944-7973},
   issue = {11},
   journal = {Water Resources Research},
   month = {11},
   pages = {9592-9603},
   publisher = {John Wiley \& Sons, Ltd},
   title = {Statistical Inference Over Persistent Homology Predicts Fluid Flow in Porous Media},
   volume = {55},
   url = {https://onlinelibrary.wiley.com/doi/full/10.1029/2019WR025171},
   year = {2019}
}

@article{Ishihara2023,
   author = {Shingo Ishihara and George Franks and Junya Kano},
   doi = {10.1016/J.APT.2022.103874},
   issn = {0921-8831},
   issue = {1},
   journal = {Advanced Powder Technology},
   month = {1},
   pages = {103874},
   publisher = {Elsevier},
   title = {Effect of particle packing structure on the elastic modulus of wet powder compacts analyzed by persistent homology},
   volume = {34},
   year = {2023}
}

@article{Jiang2018,
   author = {Fei Jiang and Takeshi Tsuji and Tomoyuki Shirai},
   doi = {10.1029/2017WR021864},
   issn = {19447973},
   issue = {6},
   journal = {Water Resources Research},
   month = {6},
   pages = {4150-4163},
   publisher = {Blackwell Publishing Ltd},
   title = {Pore Geometry Characterization by Persistent Homology Theory},
   volume = {54},
   year = {2018}
}

@article{Robins2016,
   author = {Vanessa Robins and Mohammad Saadatfar and Olaf Delgado-Friedrichs and Adrian P. Sheppard},
   doi = {10.1002/2015WR017937},
   issn = {19447973},
   issue = {1},
   journal = {Water Resources Research},
   month = {1},
   pages = {315-329},
   publisher = {Blackwell Publishing Ltd},
   title = {Percolating length scales from topological persistence analysis of micro-CT images of porous materials},
   volume = {52},
   year = {2016}
}

@article{Lee2017,
   author = {Yongjin Lee and Senja D. Barthel and Paweł Dłotko and S. Mohamad Moosavi and Kathryn Hess and Berend Smit},
   doi = {10.1038/ncomms15396},
   issn = {2041-1723},
   issue = {1},
   journal = {Nature Communications 2017 8:1},
   month = {5},
   pages = {1-8},
   pmid = {28534490},
   publisher = {Nature Publishing Group},
   title = {Quantifying similarity of pore-geometry in nanoporous materials},
   volume = {8},
   url = {https://www.nature.com/articles/ncomms15396},
   year = {2017}
}

@article{Wang2025,
   author = {Bingxu Wang and Bin Feng and Linpeng Lv and Shunning Li and Feng Pan},
   doi = {10.1021/ACS.JPCLETT.5C01831},
   issn = {19487185},
   issue = {32},
   journal = {The Journal of Physical Chemistry Letters},
   month = {8},
   pages = {8056-8067},
   pmid = {40742172},
   publisher = {American Chemical Society},
   title = {Structural Feature Extraction via Topological Data Analysis},
   volume = {16},
   url = {https://pubs.acs.org/doi/abs/10.1021/acs.jpclett.5c01831},
   year = {2025}
}

@article{Chen2025,
   author = {Dong Chen and Chun-Long Chen and Guo-Wei Wei},
   journal={Journal of Materials Chemistry A},
   doi = {10.1039/d4ta08877h},
   title = {Category-specific topological learning of metal-organic frameworks},
   url = {https://doi.org/10.1039/d4ta08877h},
   year = {2025}
}

@article{Krishnapriyan2021,
   author = {Aditi S Krishnapriyan and Joseph Montoya and Maciej Haranczyk and Jens Hummelshøj and Dmitriy Morozov},
   doi = {10.1038/s41598-021-88027-8},
   isbn = {0123456789},
   journal = {Scientific Reports},
   pages = {8888},
   title = {Machine learning with persistent homology and chemical word embeddings improves prediction accuracy and interpretability in metal-organic frameworks},
   volume = {11},
   url = {https://doi.org/10.1038/s41598-021-88027-8},
   year = {2021}
}

@article{Herring2019,
   author = {A. L. Herring and V. Robins and A. P. Sheppard},
   doi = {10.1029/2018WR022780},
   issn = {19447973},
   issue = {1},
   journal = {Water Resources Research},
   month = {1},
   pages = {555-573},
   publisher = {Blackwell Publishing Ltd},
   title = {Topological Persistence for Relating Microstructure and Capillary Fluid Trapping in Sandstones},
   volume = {55},
   year = {2019}
}

@article{Schaedler2016,
   author = {Tobias A. Schaedler and William B. Carter},
   doi = {10.1146/annurev-matsci-070115-031624},
   issn = {15317331},
   journal = {Annual Review of Materials Research},
   month = {7},
   pages = {187-210},
   publisher = {Annual Reviews Inc.},
   title = {Architected Cellular Materials},
   volume = {46},
   year = {2016}
}

@article{JMLR:v18:16-337,
  author  = {Henry Adams and Tegan Emerson and Michael Kirby and Rachel Neville and Chris Peterson and Patrick Shipman and Sofya Chepushtanova and Eric Hanson and Francis Motta and Lori Ziegelmeier},
  title   = {Persistence Images: A Stable Vector Representation of Persistent Homology},
  journal = {Journal of Machine Learning Research},
  year    = {2017},
  volume  = {18},
  number  = {8},
  pages   = {1--35},
  url     = {http://jmlr.org/papers/v18/16-337.html}
}

@book{Nye1957,
  author    = {Nye, J. F.},
  title     = {Physical Properties of Crystals: Their Representation by Tensors and Matrices},
  publisher = {Oxford University Press},
  address   = {Oxford},
  year      = {1957}
}

@book{Torquato2002,
  author    = {Torquato, Salvatore},
  title     = {Random Heterogeneous Materials: Microstructure and Macroscopic Properties},
  series    = {Interdisciplinary Applied Mathematics},
  volume    = {16},
  publisher = {Springer},
  address   = {New York},
  year      = {2002}
}

@article{Kanatani1984,
  author  = {Kanatani, Ken-Ichi},
  title   = {Distribution of directional data and fabric tensors},
  journal = {International Journal of Engineering Science},
  volume  = {22},
  number  = {2},
  pages   = {149--164},
  year    = {1984}
}

@article{Cowin1985,
  author  = {Cowin, Stephen C.},
  title   = {The relationship between the elasticity tensor and the fabric tensor},
  journal = {Mechanics of Materials},
  volume  = {4},
  number  = {2},
  pages   = {137--147},
  year    = {1985}
}

@book{montgomery2018,
  author    = {Montgomery, Douglas C. and Runger, George C.},
  title     = {Applied Statistics and Probability for Engineers},
  edition   = {7},
  publisher = {Wiley},
  address   = {Hoboken, NJ},
  year      = {2018}
}

@book{devore2020,
  author    = {Devore, Jay L.},
  title     = {Probability and Statistics for Engineering and the Sciences},
  edition   = {9},
  publisher = {Cengage Learning},
  address   = {Boston, MA},
  year      = {2020}
}

@article{Michel2001,
author = {Michel, J. C. and Moulinec, H. and Suquet, P.},
title = {A computational scheme for linear and non-linear composites with arbitrary phase contrast},
journal = {International Journal for Numerical Methods in Engineering},
volume = {52},
number = {1-2},
pages = {139-160},
doi = {https://doi.org/10.1002/nme.275},
url = {https://onlinelibrary.wiley.com/doi/abs/10.1002/nme.275},
eprint = {https://onlinelibrary.wiley.com/doi/pdf/10.1002/nme.275},
year = {2001}
}

@inproceedings{Ahrens2005ParaViewAE,
  title={ParaView: An End-User Tool for Large-Data Visualization},
  author={James Paul Ahrens and Berk Geveci},
  booktitle={The Visualization Handbook},
  year={2005},
  url={https://api.semanticscholar.org/CorpusID:56558637}
}

@inproceedings{He2015,                                                           title     = {Delving Deep into Rectifiers: Surpassing Human-Level Performance on ImageNet Classification},
author    = {He, Kaiming and Zhang, Xiangyu and Ren, Shaoqing and Sun, Jian},      
booktitle = {Proceedings of the IEEE International Conference on Computer Vision (ICCV)},
pages     = {1026--1034},                                                          
year      = {2015},                                                                
doi       = {10.1109/ICCV.2015.123}                                                
}

\end{document}